\documentclass[aps,prd,nofootinbib,superscriptaddress,preprint,amssymb,12 pt]{revtex4-1}
\usepackage{cancel}
\usepackage{amsfonts}
\usepackage{amsmath}
\usepackage{booktabs}
\usepackage{multirow}
\usepackage{slashed}
\usepackage{graphicx,color}
\usepackage{subfigure}
\usepackage{enumerate}
\usepackage[
            citecolor=blue,
            pdfstartview=FitH,
            bookmarksnumbered=true,
            bookmarksopen=true,
            anchorcolor=blue,
            menucolor=red,
            linkcolor=blue,
            filecolor=red,
            runcolor=red,
            urlcolor=blue,
            frenchlinks=red
            ]{hyperref}
\usepackage{indentfirst}

\hfuzz=\maxdimen
\begin{document}

\title{Higgs production in association with bottom quark pair at LHC}

\author{Wen-Tao Huang}
\affiliation{School of Physics and Technology, University of Jinan, Jinan Shandong 250022,  China}
\author{Hong-Lei Li}\email{sps$_$lihl@ujn.edu.cn}
\affiliation{School of Physics and Technology, University of Jinan, Jinan Shandong 250022,  China}
\author{Shi-Yuan Li}\email{lishy@sdu.edu.cn}
\affiliation{School of Physics, Shandong University, Jinan, Shandong 250100,  China}
\author{Peng-Cheng Lu}\email{pclu@mail.sdu.edu.cn}
\affiliation{School of Physics, Shandong University, Jinan, Shandong 250100,  China}
\author{Zong-Guo Si}\email{zgsi@sdu.edu.cn}
\affiliation{School of Physics, Shandong University, Jinan, Shandong 250100,  China}
\affiliation{CCEPP,IHEP,Beijing 100049, China}
\author{Ying Wang}\email{wangy@ucas.ac.cn}
\affiliation{School of Physical Sciences, University of Chinese Academy of Sciences, YuQuan Road 19A, Beijing 100049, China}
\author{Zhong-Juan Yang}\email{sps$_$yangzj@ujn.edu.cn}
\affiliation{School of Physics and Technology, University of Jinan, Jinan Shandong 250022,  China}

\begin{abstract}
The direct observation of the Yukawa interaction related to Standard Model Higgs and the bottom quark is obtained from the $h\to b\bar{b}$ process recently. The b-tagging becomes an important tool to search for new heavy resonance production at LHC. We use the b-tagging method to investigate the neutral Higgs production in association with a bottom quark pair, and perform the detector simulation for the signal process with the Higgs decaying into $b\bar{b}$ or $\tau^+\tau^-$ together with the dominant backgrounds. Our results show that  $b\bar{b}b\bar{b}$ and $b\bar{b}\tau^+\tau^-$ processes will play an important role to search for the exotic heavy Higgs production at LHC and future hadron colliders.
\end{abstract}

\maketitle

\section{Introduction}
The discovery of the Higgs boson at the Large Hadron Collider (LHC) confirms the importance to search for the exotic Higgs bosons beyond Standard Model (SM). There are several well-known motivations for the extension of Higgs sectors, such as CP-violation~\cite{Cheng:1987gp,Gudkov:1991qg}, Supersymmetry~\cite{Nilles:1983ge,Fayet:1976cr,Sohnius:1985qm}, Grand Unification~\cite{Georgi:1974sy,Georgi:1974yf}, etc. One simple example is the Two-Higgs-Doublet Model (2HDM)~\cite{Ellis:1985xy,Weinberg:1976hu,Liu:1987ng}, where there are three neutral Higgs bosons (h, H, A) and two charged Higgs bosons ($H^{\pm}$). In the 2HDM, flavour changing neutral currents at tree level can be suppressed by the discrete symmetries, thus the fermions can be coupled to the Higgs doublets in different types. The model with up-type quarks coupling to one Higgs doublet while down-type quarks and charged leptons coupling to the other one is called type-II 2HDM, where the Higgs sector is the same as that in Supersymmetry models. This kind of model provides abundant phenomena to search for new physics related to exotic heavy Higgs. The LHC have performed extensive 2HDM interpretations of measurement via various channels. The direct searches for neutral heavy Higgs bosons include the decay channel $\tau^+\tau^-$~\cite{Aaboud:2017sjh,CMS:2017epy}, $WW/ZZ$~\cite{Aaboud:2017gsl,Aaboud:2017rel,Sirunyan:2018qlb}, $\gamma\gamma$~\cite{Aaboud:2017yyg}, $A\to hZ$~\cite{Aaboud:2017cxo}, $A/H\to HZ/AZ$~\cite{Aaboud:2018eoy,Khachatryan:2016are}, and $H\to hh$~\cite{Aaboud:2018ftw,Sirunyan:2018iwt}. The strongest bounds at large $\tan\beta$ come from $A/H\to \tau^+\tau^-$ mode, which exclude $m_{A/H}$ about 1.5 TeV for $\tan\beta\sim 50$~\cite{Chen:2018shg}. In the small $\tan\beta$ region, the $A/H\to tt$ mode can help to extend the exclusion limit up to 2 TeV~\cite{Craig:2016ygr}. The direct searches for charged Higgs bosons are reported with the $H^\pm\to \tau\nu,tb$ channels~\cite{ATLAS:2016qiq,Aaboud:2018gjj,CMS:2016szv}. The D0 and CDF Collaboration have presented a search for $H\to b\bar{b}$, produced in association with bottom quarks~\cite{Aaltonen:2011nh,Abazov:2005yr,Abazov:2008hh}. The theoretical  constraints on the 2HDM parameters are reviewed with the vacuum stability, unitarity, electroweak precision measurements and flavour constraints~\cite{Kling:2016opi}.

The Yukawa coupling is proportional to the fermion mass. As a result, it is promising to investigate the Yukawa interaction between Higgs boson and the third generation fermions ($t, b, \tau$). The Higgs boson production in association with a top quark pair process is widely studied theoretically~\cite{Li:2015kaa,Cao:2016wib,Bernreuther:2017yhg,Bernreuther:1997gs,Bernreuther:2016nuv} and is observed at LHC for the first time~\cite{Sirunyan:2018hoz,Aaboud:2018urx}. In the SM, Higgs production in associated with bottom quark is suppressed due to small Yukawa coupling compared with top quark. To study the Yukawa coupling $y_{b\bar{b}h}$ related to Higgs and bottom quark, the $b\bar{b}h$ production with $h\to b\bar{b}$ process is interesting, since there is only one type Yukawa interaction. The inclusive $b\bar{b}h$ cross section can be enhanced by the QCD effect and phase space~\cite{Dawson:2003kb,Dittmaier:2003ej}. Moreover, in the 2HDM with a large value of $\tan\beta$, the cross section of $b\bar bH$ (A) can be larger than that of the $t\bar tH$ (A) process. Recently, the direct observation of the Yukawa interaction related to SM Higgs and the bottom quark is obtained from the $h\to b\bar{b}$ process at LHC~\cite{Aaboud:2018zhk,Sirunyan:2018kst}. Comparing the $b\bar{b}$ decay mode of Higgs, though the branching ratio of $\tau^+\tau^-$ decay mode is not very large, it can also provide significant signatures for the Higgs production due to the clear background. The b-tagging method is well developed and becomes an important tool to search for new heavy resonance production at LHC. In this work, as an example, we use the b-tagging method to investigate the neutral Higgs production in association with a bottom quark pair via the signature of $b\bar{b}b\bar{b}$ and $b\bar{b}\tau^+\tau^-$ within the 2HDM framework at LHC. The corresponding background processes such as $bbjj$, $Zbb$, $ZZ$ at LHC are also simulated.

This paper is organized as follows. In Sec.\ref{Sec:framework}, the corresponding theoretical framework is briefly introduced. The numerical results with $h/H/A\rightarrow b\bar{b}/\tau^+\tau^-$ decay modes in the SM and 2HDM are listed in Sec.\ref{Sec:collider}. Finally, a brief summary is given.
\section{Theoretical Framework}\label{Sec:framework}
\subsection{Two-Higgs-Doublet Model}
In the Standard Model, there is one $\mathrm{SU}(2)$ scalar doublet. After spontaneous symmetry breaking, it is well known that the mass of fermion is extracted from the related Yukawa interactions,
 \begin{align}
 \mathcal{L}^{SM}_{Yukawa}=-\frac{m_f}{v}f\bar{f}H,
 \end{align}
 where $m_f$ is the mass of fermion and $v$ is the vacuum expectation value of Higgs. It shows the interaction of Higgs boson coupling to bottom quarks is proportional to the mass of bottom quark. However, this Yukawa coupling could be considerably enhanced in the extensions of the standard model with more than one Higgs doublet. The two-Higgs-doublet model is the simplest extension to the SM, which contains two scalar Higgs doublets $\Phi_1$ and $\Phi_2$. In the 2HDM, there are two neutral, scalar Higgs bosons. One of them can be taken as the SM-like Higgs boson which has already been discovered at the LHC. In addition, there are two charged Higgs bosons  as well as a neutral, pseudoscalar Higgs boson.

 In the conventions of references~\cite{Davidson:2005cw,Bernreuther:2015fts}, the most general gauge invariant and renormalizable potential that can be formed using two Higgs doublets is given by
 \begin{align}
 V(\Phi_1,\Phi_2)=&m_{11}^2\Phi_1^\dag\Phi_1+m_{22}^2\Phi_2^\dag\Phi_2-[m_{12}^2\Phi_1^\dag\Phi_2+h.c.]\nonumber\\
 &+\frac{\lambda_1}{2}(\Phi_1^\dag\Phi_1)^2+\frac{\lambda_2}{2}(\Phi_2^\dag\Phi_2)^2\nonumber\\
 &+\lambda_3(\Phi_1^\dag\Phi_1)(\Phi_2^\dag\Phi_2)+\lambda_4(\Phi_1^\dag\Phi_2)(\Phi_2^\dag\Phi_1)\nonumber\\
 &+{\frac{\lambda_5}{2}(\Phi_1^\dag\Phi_2)^2+[\lambda_6(\Phi_1^\dag\Phi_1)+\lambda_7(\Phi_2^\dag\Phi_2)](\Phi_1^\dag\Phi_2)+h.c.},
 \label{eq:higgs potential}
 \end{align}
where $m_{11}^2, m_{22}^2$ and $\lambda_1,...,\lambda_4$ are real parameters, In general, $m_{12}^2, \lambda_5, \lambda_6$ and $\lambda_7$ are complex. This Higgs potential violates the CP symmetry while the CP-conserving Higgs potential may be motivated by assuming an approximate $Z_2$ symmetry. This approximate symmetry implies that the hard $Z_2$ symmetry-breaking terms in \eqref{eq:higgs potential} are absent, $\lambda_6=\lambda_7=0$. After electroweak gauge symmetry breaking, there are two unitary gauge $\mathrm{SU}(2)$ doublets with the formula of
\begin{align}
\Phi_1=\left(\begin{array}{c}
-H^+\sin\beta,\\
(v_1+\varphi_1-iA\sin\beta)/\sqrt{2}
\end{array}\right),&~~~\Phi_2=\left(\begin{array}{c}
H^+\cos\beta,\\
(v_2+\varphi_2+iA\cos\beta)/\sqrt{2}
\end{array}\right),
\end{align}
where $v_1, v_2$ are the vacuum expectation values with $v=\sqrt{v_1^2+v_2^2}=246~\mathrm{GeV}$, and $\tan\beta=v_2/v_1$ is the ratio of the vacuum expectation values of the two-Higgs-doublet fields. The field $H^+$ describes the physical charged Higgs boson of the model, while $\varphi_{1,2}$ and A are the physical neutral CP-even and CP-odd fields. We denote the mass eigenstates of these neutral Higgs fields as $\phi_k$ ($k=1,2,3$), which are related with the physical states by an orthogonal transformation $(\phi_1,\phi_2,\phi_3)^T = \mathrm{R}(\varphi_1,\varphi_2,A)$. From here on, $k=1,2,3$ if there is no additional explicit definition. The orthogonal  matrix $\mathrm{R}$ can be parameterized as follows,
\begin{align}
\mathrm{R}=\left(\begin{array}{ccc}
-s_1c_2                         &     c_1c_2                       &s_2  \\
s_1s_2s_3-c_1c_3     &     -s_1c_3-c_1s_2s_3     &c_2s_3\\
s_1s_2c_3+c_1s_3    &     s_1s_3-c_1s_2c_3    &c_2c_3
\end{array}\right),
\label{R marix}
\end{align}
where $c_i$ and $s_i$ ($i=1,2,3$) are the cosines and sines of the mixing angles $\alpha_i$. In the case of CP-conserving, we can require only $\phi_1$ and $\phi_2$ mixing, which means $\alpha_1\neq0,~\alpha_2=\alpha_3=0$. The matrix R is now block diagonal with $R_{13}=R_{23}=R_{31}=R_{32}=0$ and $R_{33}=1$. The three neutral Higgs conventionally are denoted by $\phi_1=h, \phi_2=H,$ and $\phi_3=A$. Then seven independent parameters are left in the mass basis,
\begin{align}
m_h,~~m_H,~~m_A,~~m_{H^\pm},~~v,~~\alpha_1,~~\tan\beta.
\end{align}
The alignment limit is supposed to be consistent with the SM Higgs boson gauge couplings with $\sin(\beta-\alpha_1)=1$~\cite{Inoue:2014nva,Chen:2015gaa}.  In the 2HDM, flavour changing neutral currents at tree level can be suppressed by the discrete symmetries, thus the fermions can be coupled to the Higgs doublets in different types. In the type-I 2HDM, all quarks and charged leptons couple to $\Phi_2$ doublet. In the type-II 2HDM, the doublet $\Phi_1$ is coupled to down-type quarks and charged leptons, while $\Phi_2$ is coupled to up-type quark. In the lepton-specific 2HDM, the doublet $\Phi_1$ is coupled to charged leptons, while $\Phi_2$ is coupled to all quarks. In the flipped 2HDM, the doublet $\Phi_1$ is coupled to down-type quarks while $\Phi_2$ is coupled to charged leptons and up-type quarks.
 We study the Higgs interactions in type-II 2HDM. The Yukawa couplings of the neutral scalars to fermions can be obtained through the interactions of $\phi_k$ with the third generation quarks,
\begin{align}
\mathcal{L}^{2HDM}_{Yukawa}=-\sum\limits_{f=t,b}\frac{m_f}{v}(c_1^{kf}\bar{f}f-c_2^{kf}\bar{f}i\gamma_5f)\phi_k,
\label{eq:lagrangian}
\end{align}
where $c_1^{kf}, c_2^{kf}$ are the mixing factors in Yukawa interactions. $c_1^{kf}=1,c_2^{kf}=0$ is corresponding to the pure scalar Higgs, $c_1^{kf}=0,c_2^{kf}=1$ is corresponding to the pure pesudoscalar Higgs. One can notice that the strength of coupling between Higgs and fermions depends on the values of $\tan\beta$ and the transition matrix elements $R_{ij}$ listed in Table {\ref{tab:couplings}}. It leads to three types of Yukawa interaction for the SM Higgs (h), heavy Higgs (H) and the pseudoscalar Higgs (A) respectively. Hence the Higgs with bottom quark interaction gets more complicated and leads to different properties for the production.
\renewcommand{\arraystretch}{1.2}
\begin{table}[!ht]
  \centering \caption{The couplings of Higgs boson $\phi_k$ to the up and down quarks in type-II 2HDM for Eq.\eqref{eq:lagrangian}. The labels t and b stand for u-type and d-type quarks.}\label{tab:couplings}
\begin{tabular}{cccc}\hline\hline
        $ c_1^{kt}$              ~~~       &$c_1^{kb}=c_1^{kl}$             ~~~     &$c_2^{kt}$                ~~~    &$c_2^{kb}=c_2^{kl}$     \\[1pt]\hline
        $R_{k2}/\sin\beta$     ~~~       &$R_{k1}/\cos\beta$          ~~~     &-$R_{k3}/\cot\beta$     ~~~    &-$R_{k3}\tan\beta$  \\\hline\hline
\end{tabular}
\end{table}
\subsection{Decay width}
The decay branching ratio of Higgs boson is model-dependent. In the SM, the decay branching ratio $\mathrm{BR}(h\to b\bar{b})$ is 60\% and $\mathrm{BR}(h\to \tau^+\tau^-$) is around 6\%~\cite{Tanabashi:2018oca}. In the 2HDM, according to the above discussion and  constraints from the type-II model with CP-conserving, the parameters are set as
\begin{align}\label{parameter1}
&\beta-\alpha_1=\pi/2, ~\alpha_2=\alpha_3=0,~v=246~\mathrm{GeV},\nonumber\\
&m_h=125~\mathrm{GeV},~m_H=m_A,~m_{H^{\pm}}=m_H+200~\mathrm{GeV},
\end{align}
where the charged Higgs is heavier than the neutral Higgs. The decay modes of the heavy Higgs (H, A) to $H^{\pm}W^{\mp}$ and $H^{\pm}H^{\mp}$ are forbidden.

\begin{figure}[!ht]
\begin{center}
\subfigure[]{ \label{fig:branchratio_tanbeta_A}
\includegraphics[scale=0.35]{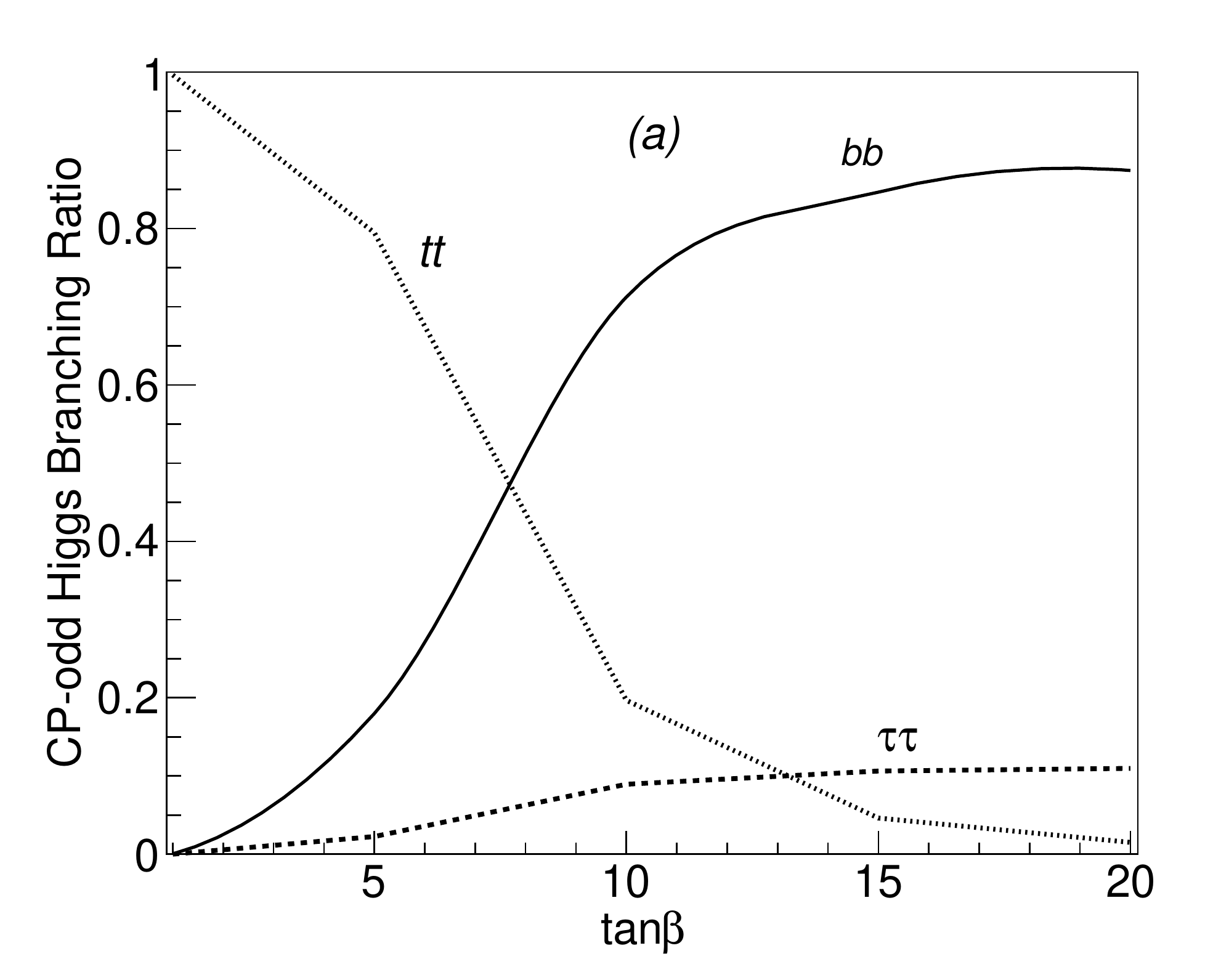}
}
\subfigure[]{ \label{fig:branchratio_tanbeta_H}
\includegraphics[scale=0.35]{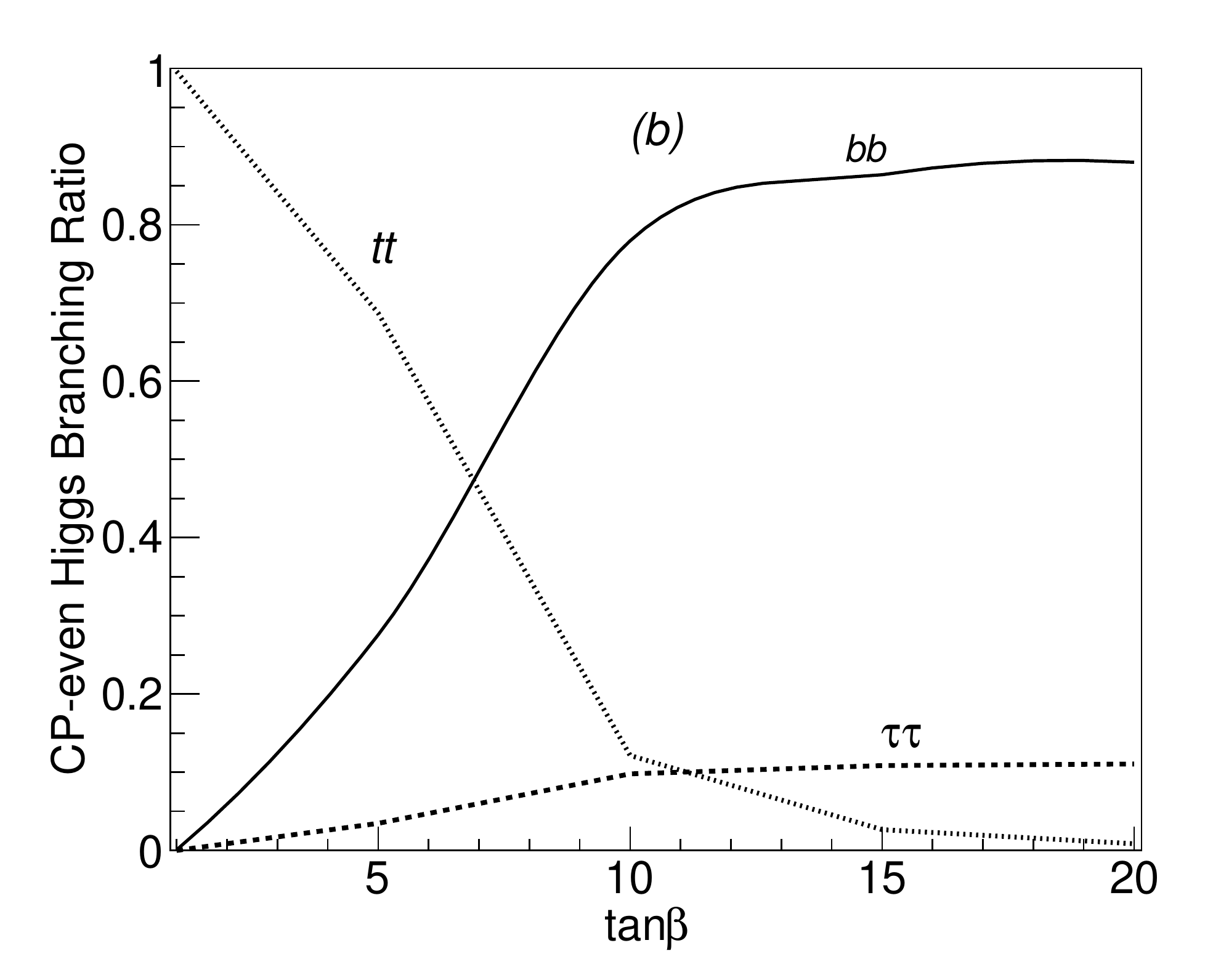}
}
\subfigure[]{ \label{fig:branchratio_2hdm_A}
\includegraphics[scale=0.35]{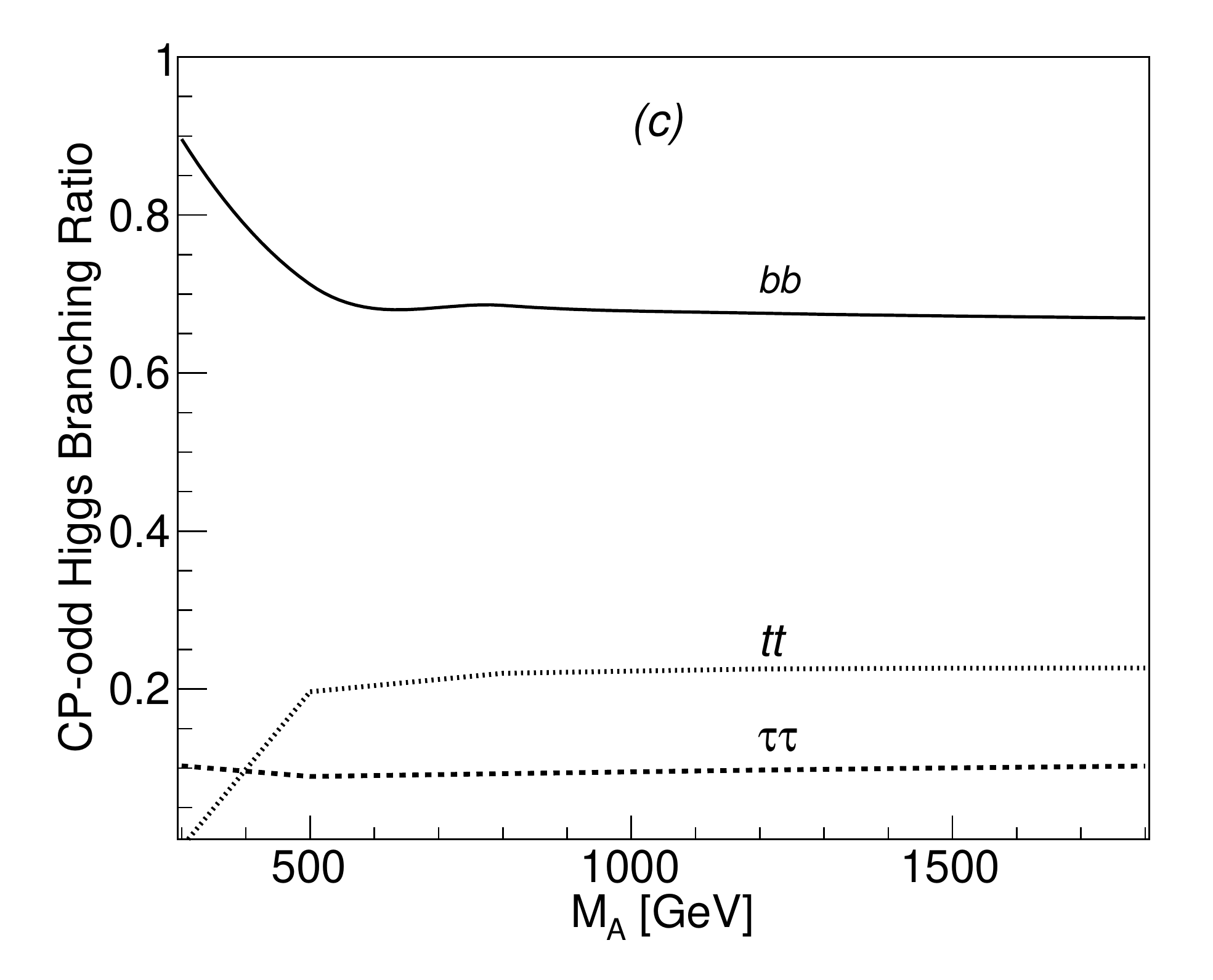}
}
\subfigure[]{ \label{fig:branchratio_2hdm_H}
\includegraphics[scale=0.35]{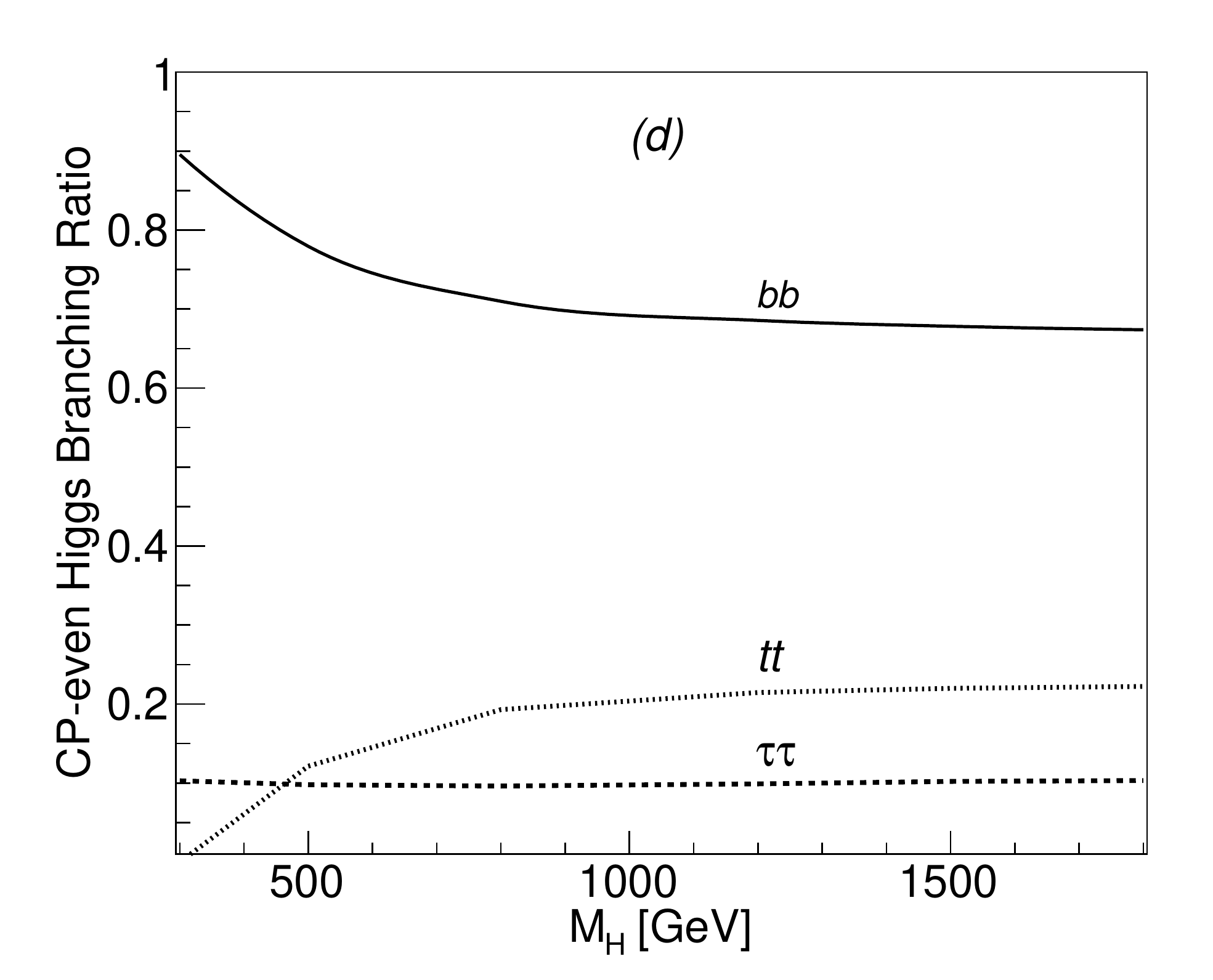}
}
\caption{The dominate decay branching ratios of the two heavy neutral Higgs bosons. (a), (b) Varying $\tan\beta$ with  $M_{H/A}=500~\mathrm{GeV}$. (c), (d) Varying the Higgs mass with $\tan\beta=10$. }\label{fig:branchratio_2hdm}
\end{center}
\end{figure}

In 2HDM, we use the 2HDMC~\cite{Eriksson:2009ws} code to calculate the Higgs boson branching ratio. The results with various mass values and $\tan\beta$ are shown in Fig.\ref{fig:branchratio_2hdm}. The dominant decay channel is $H~(A) \to b\bar{b}$ mode with $\tan\beta>10$. Its branching ratio is about eighty percent. The branching ratio of $\tau^+\tau^-$ is about ten percent.
In the following works, we analyse scalar and pesudoscalar Higgs boson production associated with bottom quark via $b\bar{b}b\bar{b},  b\bar{b}\tau^+\tau^-$ final states.
\subsection{$b\bar{b}\phi_k$ associated production at the LHC}
\begin{figure}[t]
\begin{center}
  \includegraphics[scale=0.6]{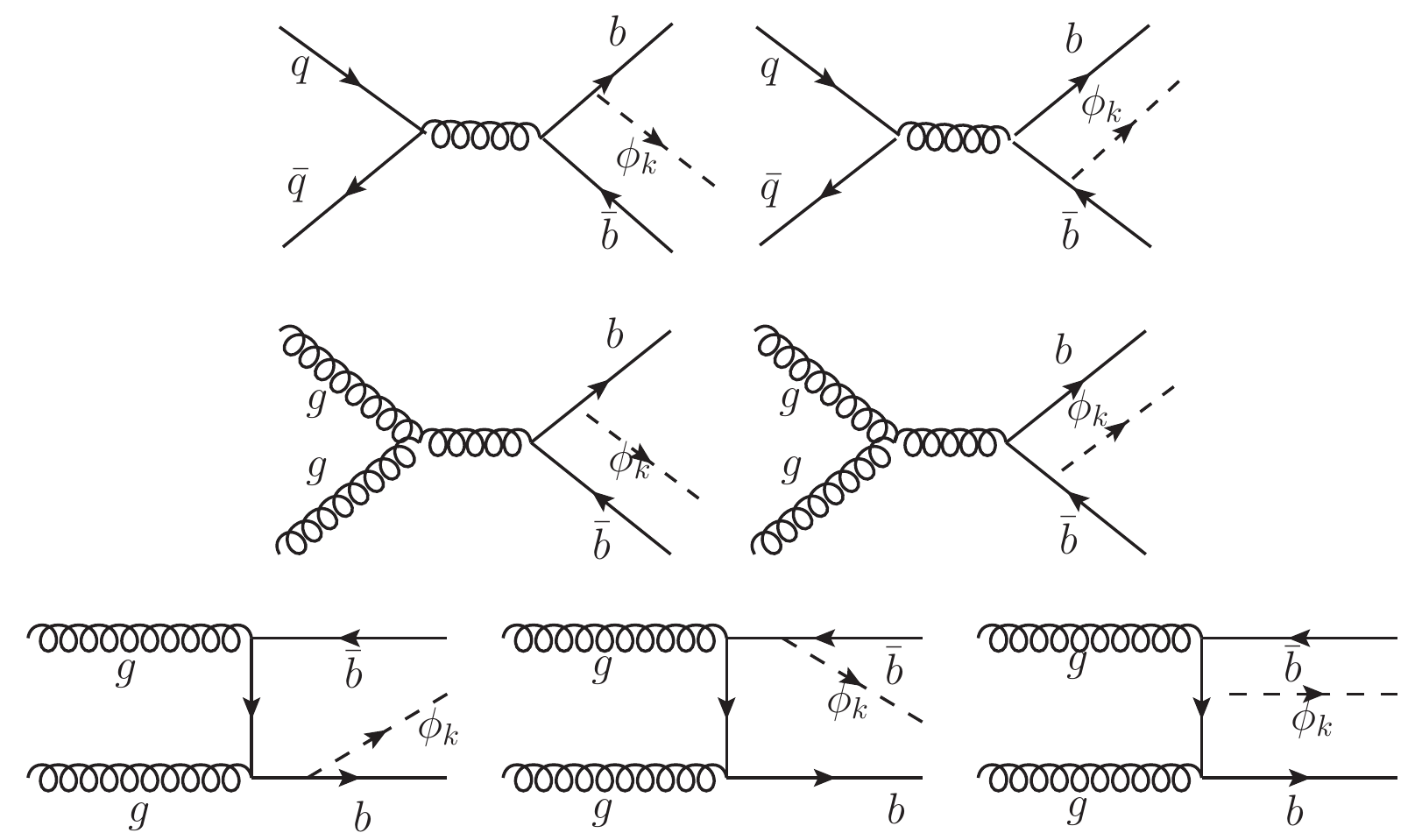}
  \caption{The Feynman diagrams for $b\bar{b}\phi_k$ production at leading order.}\label{fig:feyman diagram}
\end{center}
\end{figure}
At the proton-proton collider, we begin to investigate the $b\bar{b}\phi_k$ production with the quark annihilation and the gluon fusion processes as displayed in Fig.\ref{fig:feyman diagram},
\begin{align}
g(p_1)~+~g(p_2)~ &\rightarrow ~ b(p_3)~+~\bar{b}(p_4)~+~\phi_k(p_5),\nonumber\\
q(p_1)~+~\bar{q}(p_2)~ &\rightarrow ~ b(p_3)~+~\bar{b}(p_4)~+~\phi_k(p_5),
\label{pp2bbh process}
\end{align}
where $p_i$ is the four-momentum of the corresponding particle. The squared matrix element for process $pp \to b\bar{b}\phi_k$ can be written as
\begin{align}
|\tilde{\mathcal{M}}|^2~=~|\tilde{\mathcal{M}}_{gg}|^2~+~|\tilde{\mathcal{M}}_{q\bar{q}}|^2,
\end{align}
where $\tilde{\mathcal{M}}_{gg}~(\tilde{\mathcal{M}}_{q\bar{q}})$ is the invariant amplitude of $gg~(q\bar{q}) \to b\bar{b}\phi_k$, the expression of $|\tilde{\mathcal{M}}_{gg}|^2$ is lengthy, and is not listed here. The expression of $|\tilde{\mathcal{M}}_{q\bar{q}}|^2$ has the formula of
\begin{align}
 |\tilde{\mathcal{M}}_{q\bar{q}\to b\bar{b}\phi_k}|^2~=~\frac{2(N^{2}_{c}-1)g_{s}^{4}}{N_{c}^{2}s^{2}}\Big\{&({c_1^{kb}}^{2}+{c_2^{kb}} ^{2})\big[\mathcal{A}_{1}~+~\mathcal{A}_{1}\mid_{p_{3}\leftrightarrow p_{4}}~+~\mathcal{A}_{2}\big]\nonumber\\
  ~+~&({c_1^{kb}}^{2}-{c_2^{kb}}^{2})\big[\mathcal{B}_{1}~+~\mathcal{B}_{1}\mid_{p_{3}\leftrightarrow p_{4}}~+~\mathcal{B}_{2}\big]\Big\}.
\end{align}
The $\mathcal{A}_{1},\mathcal{A}_{2},\mathcal{B}_{1}$ and $\mathcal{B}_{2}$ are expressed as
\begin{align}
\mathcal{A}_1=&\frac{1}{s_2^2}\Big[s_2(t_2u_1+t_1u_2)+(t_1-s_1+u_1)(t_2(s-4u_2)+s(m^2+u_2))\Big],\\
\mathcal{A}_2=&\frac{1}{s_2s_3}\Big[2t_2^2u_1+t_2(m^2s-s\cdot s_1+2u_1(-2s_1+u_1-u_2))+2t_1^2u_2+t_1(m^2s-s\cdot s_1+2t_2\nonumber\\
&(s-u_1-u_2)-4s_1u_2-2u_1u_2+2u_2^2)+s(2s_1^2+2u_1u_2+m^2(u_1+u_2)-s_1(u_1+u_2))\Big],\\
 \mathcal{B}_1=&\frac{1}{2s_2^2}\Big[m^2(2m^2s+s^2-8t_2u_2)\Big],\\
 \mathcal{B}_2=&\frac{1}{s_2s_3}\Big[m^2(s^2-2s\cdot s_1+4t_2u_1+4t_1u_2)\Big],
\end{align}
where $s$, $s_i$ ($i=1,2,3$), $t_i$, $u_i$ ($i=1,2$) are defined as
\begin{align}
&s=(p_1+p_2)^2, ~~s_1=p_3\cdot p_4, ~~s_2=m^2-(p_3+p_5)^2, ~~s_3=m^2-(p_4+p_5)^2,\nonumber\\
&t_1=p_1\cdot p_3,~~t_2=p_1\cdot p_4,~~u_1=p_2\cdot p_3,~~u_2=p_1\cdot p_4,
\end{align}
with the bottom mass of $m$.

The cross section with the help of parton distribution functions for $pp \to b\bar{b}\phi_k$ can be written as:
 \begin{align}
 \sigma(pp \to b\bar{b}\phi_k)=\int dx_1dx_2[&(f_{q/h_1}(x_1) f_{\bar{q}/h_2}(x_2)+f_{q/h_2}(x_2) f_{\bar{q}/h_1}(x_1))\hat{\sigma}_{q\bar{q}}\nonumber\\
 &+f_{g/h_1}(x_1) f_{g/h_2}(x_2)\hat{\sigma}_{gg}].
 \end{align}
Given the Higgs boson is produced on-shell, with the narrow width approximation, we have
\begin{align}
   \lim\limits_{\Gamma\rightarrow 0} \frac{1}{(p_5^2-M^2)^2+\Gamma^2M^2} \rightarrow \frac{\pi}{\Gamma M}\delta(p_5^2-M^2),
\end{align}
where $\Gamma$ and $M$ denote the total decay width of Higgs boson and mass respectively. The total cross section including Higgs decay information can be written as
\begin{align}
\sigma &= \sigma(pp \to b\bar{b}\phi_k)~\times~ \mathrm{BR}(\phi_k \to f\bar f).
\end{align}
\section{Collider Analysis}\label{Sec:collider}
\begin{figure}[!ht]
\begin{center}
\subfigure[]{
\includegraphics[scale=0.35]{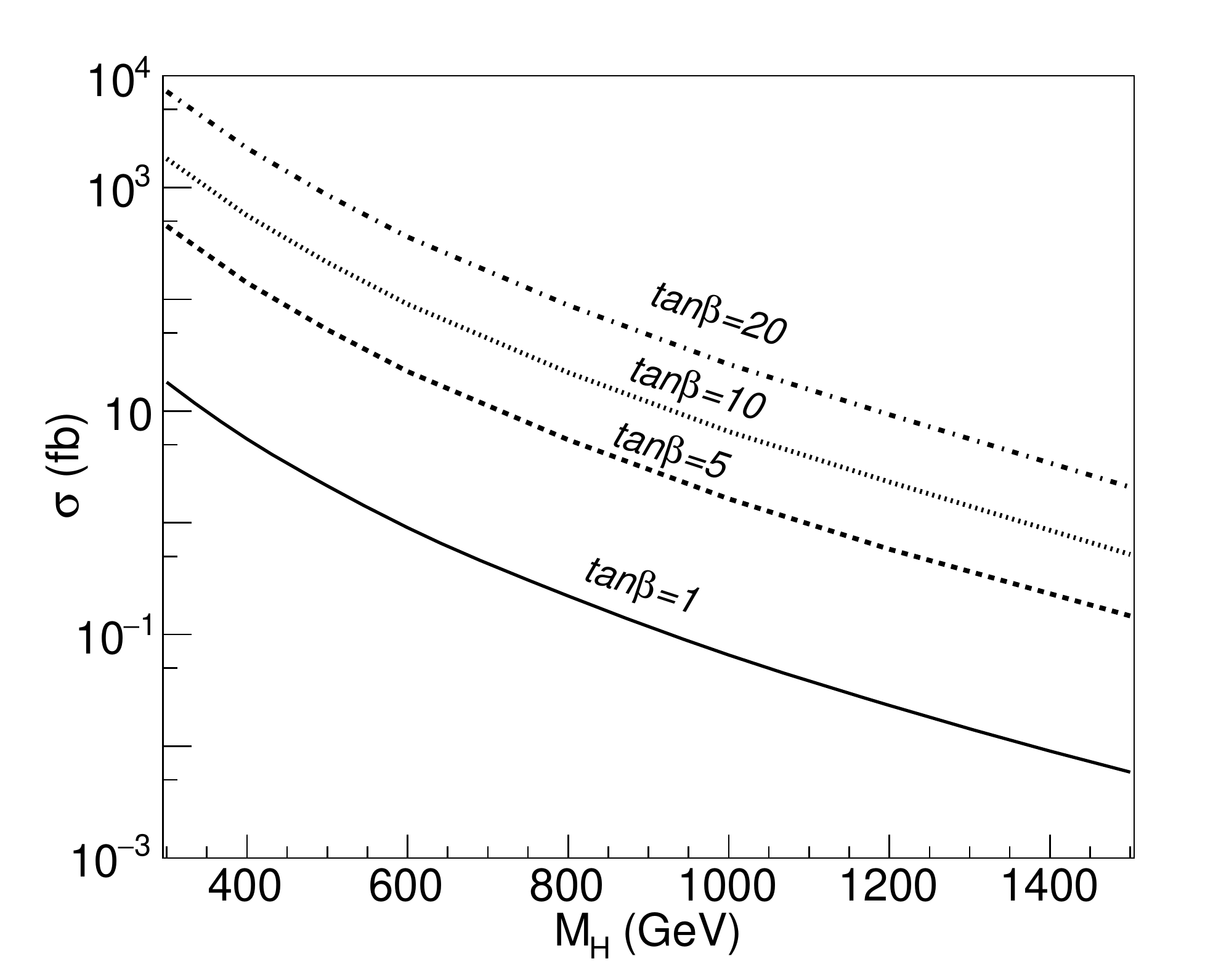}
}
\subfigure[]{
\includegraphics[scale=0.35]{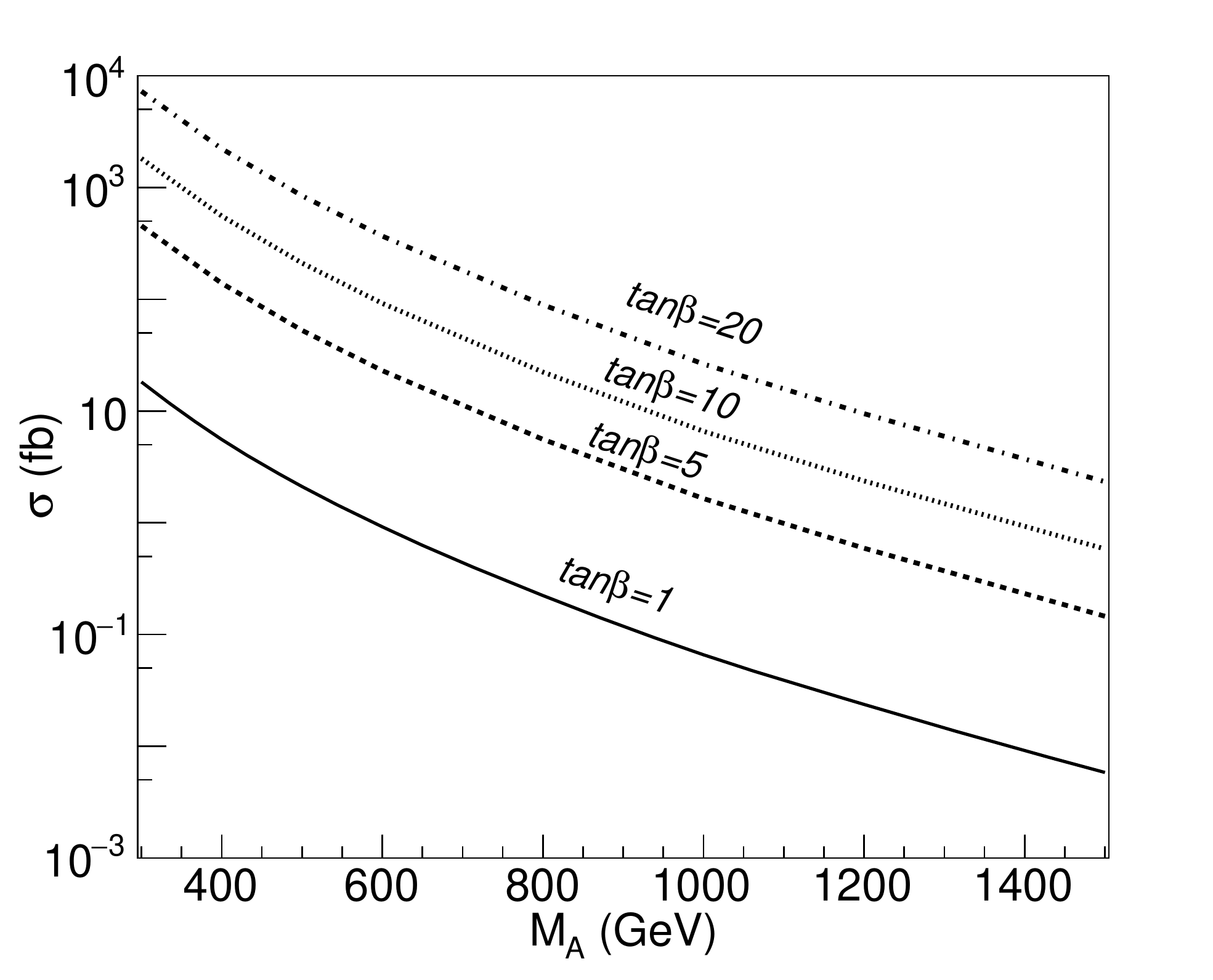}
}
\caption{The cross section for signal process $pp \to b\bar{b}H/A$ at the LHC $14~\mathrm{TeV}$.}\label{fig:pp2bbhcs_tanbeta_m}
\end{center}
\end{figure}

 The cross sections for $b\bar{b}H$ and $b\bar{b}A$ production at LHC 14~$\mathrm{TeV}$ are ploted as a function of Higgs boson mass in Fig.\ref{fig:pp2bbhcs_tanbeta_m}. With the integrated luminosity of 300 $\mathrm{fb}^{-1}$, one can find that there are significant number of events for $b\bar{b}H$ and $b\bar{b}A$ production for $m_{A/H}<1.4~\mathrm{TeV}$ and $\tan\beta \geq 1$, and there is no significant discrepancy between these two processes.

  The SM Higgs boson and exotic Higgs can be produced in the same process and lead to the same collider signatures. We investigate the neutral Higgs production in association with bottom quark pair with  $\phi_k\to b\bar{b}$ and $\phi_k\to \tau^+\tau^-$ modes,
\begin{align}
pp &\to b\bar{b}\phi_k \to b\bar{b}b\bar{b}, \label{process_ppt4b} \\
pp &\to b\bar{b}\phi_k \to b\bar{b}\tau^+\tau^-. \label{process_ppt2b2tau}
\end{align}
  Our numerical results are obtained with CTEQ6L1 parton distribution function. We choose the bottom quark mass $m=4.8~\mathrm{GeV}$ and $\tau$ lepton mass $m_{\tau}=1.776~\mathrm{GeV}$. To be more realistic, the simulation at the detector is performed by smearing the lepton and jet energies according to the assumption of the Gaussian resolution parametrization
 \begin{align}\label{Gaussian}
   \frac{\delta(E)}{E}=\frac{a}{\sqrt{E}}\oplus b,
\end{align}
 where $\delta(E)/E$ is the energy resolution, $a$ is a sampling term, $b$ is a constant term, and $\oplus$ denotes a sum in quadrature. We take $a=5\%$, $b=0.55\%$ for leptons and $a=100\%$, $b=5\%$ for jets, respectively~\cite{Aad:2009wy}.

 After smearing the energy of jets and leptons, we use the following strategy to confirm the $b\bar{b} b\bar{b}$ signal events. The b-tagging efficiency is supposed to be 60\%. Firstly, the events with four jets are chosen. Then two of the four jets are chosen optionally to reconstruct the invariant mass. The two jets that  reconstructed mass closest to the Higgs mass are defined as $j_1$ and $j_2$ and the remaining two are labeled as $j_3$ and $j_4$. We require the transverse momentum $p_T^{j_1}>p_T^{j_2}$ and $p_T^{j_3}>p_T^{j_4}$. For the  $b\bar{b}\tau^+\tau^-$ final state, we use the invariant mass of $\tau^+\tau^-$ to reconstruct the resonance. In order to identify the isolated particle, the angular distribution between particle $i$ and particle $j$ is define by
  \begin{align}\label{R}
   \Delta R_{ij} &= \sqrt{\Delta \phi^2_{ij}+\Delta\eta^2_{ij}}~,
 \end{align}
  where $\Delta\phi_{ij}$ ($\Delta \eta_{ij}$) denotes azimuthal angle (rapidity) difference between the particles. The transverse momentum $p_T$ and rapidity $\eta_j$ of the final particles are set the minimum values for the basic cuts. We simulate the detector acceptance by requiring all final states to have $p_T > 20~\mathrm{GeV}, ~|\eta| < 3$, and to be separated by a cone of $\Delta R > 0.4$.
 \subsection{SM Higgs Results}\label{sec.sm}
\begin{figure}[!ht]
   \centering
   \includegraphics[scale=0.5]{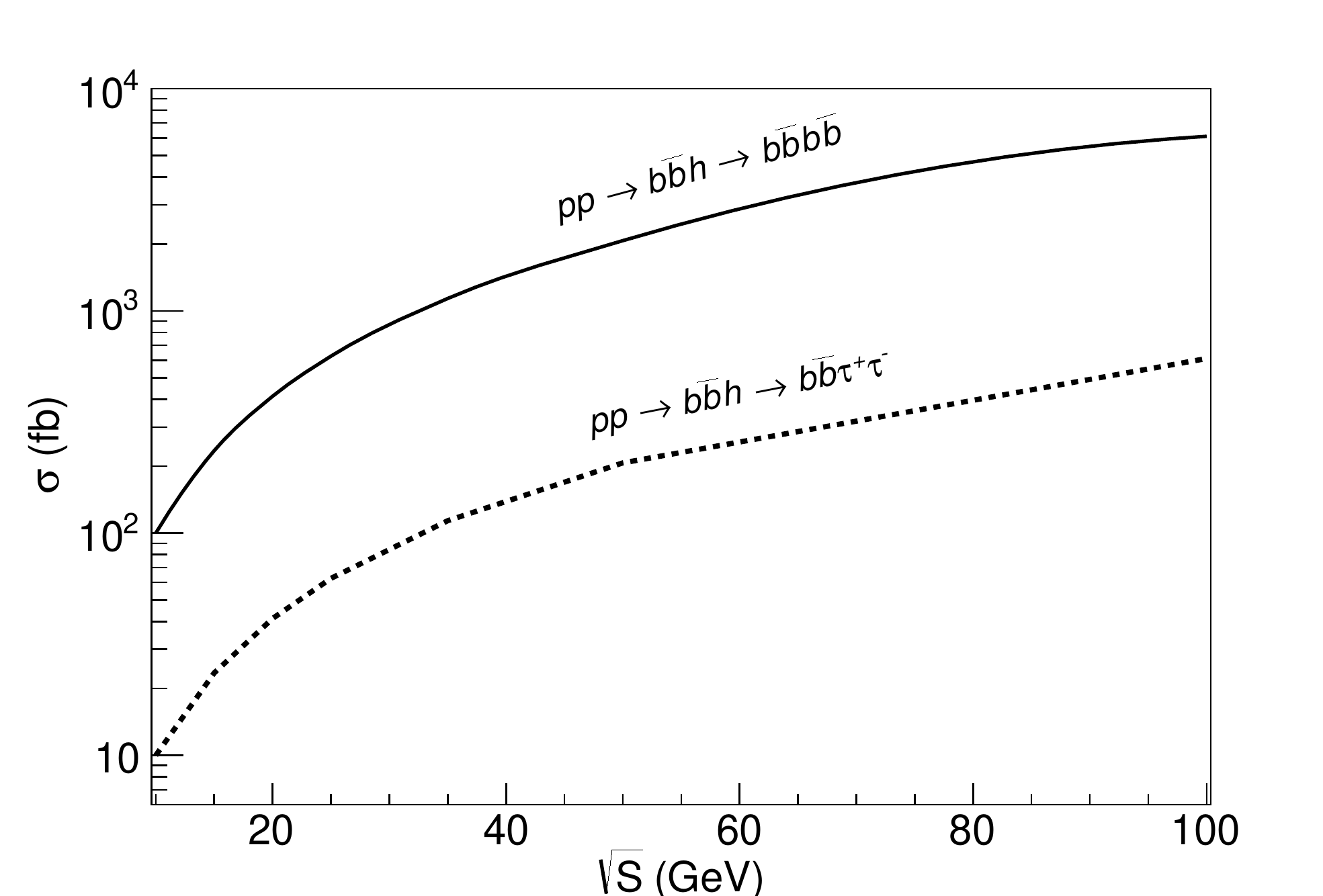}
   \caption{The cross sections as a function of center of mass energy for the processes $pp \to b\bar{b}h \to b\bar{b}b\bar{b}$ and $pp \to b\bar{b}h \to b\bar{b}\tau^+\tau^-$ respectively.}
   \label{fig:s_cs_insm}
\end{figure}

 In this section, we study the associated production of $b\bar{b}$ and SM Higgs (h) with $h\to b\bar{b}$ and $\tau^+\tau^-$. The total cross sections are shown in Fig.\ref{fig:s_cs_insm} without cuts at the LHC with different center-of-mass energy. The cross section can reach $200~\rm{fb}$ ($20~\rm{fb}$) for $b\bar{b}b\bar{b}$ ($b\bar{b}\tau^+\tau^-$) with $\sqrt{s}=14~\mathrm{TeV}$, which can be up to $6000~\rm{fb}$ ($600~\rm{fb}$) with $\sqrt{s}=100~\mathrm{TeV}$ respectively. To clarify the signal from the background, we test different kinematical cuts, and adopt the efficient cuts as follows. The SM Higgs mass is only $125~\mathrm{GeV}$, comparing to the system energy between signal and background, we require the total energy of the system is about up to $300~\mathrm{GeV}$ (referred to as $M_{j_1j_2j_3j_4} > 300 ~\mathrm{GeV}$ cut). Moreover, $j_1,j_2$ with invariant mass that best reconstructs $M_h$ can reject the event if the resulting invariant mass is more or less than $\Delta M_h$ from $M_h$, where $\Delta M_h = 0.1 M_h$ is the maximum of our estimation of the experimental mass resolution (referred to as $|M_{j_1j_2}-M_h| < 0.1M_h$ cut). As for $b\bar{b}\tau^+\tau^-$ final state, we choose $\tau^+\tau^-$ to reconstruct $M_h$. The dominant backgrounds from SM are $bbjj$, $Zbb$, $ZZ$ corresponding to $b\bar{b}b\bar{b}$ and $Zbb$ corresponding to $b\bar{b}\tau^+\tau^-$ which are simulated by MADGRAPH~\cite{Alwall:2014hca} with the default sets. To reduce the intermediate Z boson backgrounds, we veto the events with the reconstructed mass of jet close to Z boson (referred to as $|M_{j_3j_4}-M_\mathrm{z}| > 20~ \mathrm{GeV}$ cut). Finally, we require that the invariant mass of $j_3,j_4$ is more than $60~\mathrm{GeV}$ (referred to as $M_{j_3j_4} > 60~ \mathrm{GeV}$ cut). This further reduces the $bbjj$ background.
 All the cuts are set as the follows, for $b\bar{b}b\bar{b}$ final ~state,
  \begin{align}
  & p_T^{j} > 20~\mathrm{GeV}, ~|\eta_j| < 3,~ \Delta R > 0.4,~ M_{j_1j_2j_3j_4} > 300 ~\mathrm{GeV},~ 3~\mathrm{b~jet~tagging},\nonumber\\
  & |M_{j_1j_2}-M_h| < 0.1M_h, ~|M_{j_3j_4}-M_\mathrm{z}| > 20~ \mathrm{GeV}, ~M_{j_3j_4} > 60~ \mathrm{GeV},
    \end{align}
    and for $b\bar{b}\tau^+\tau^-$ final~state,
 \begin{align}
  &p_T^{j} > 20~\mathrm{GeV}, p_T^{l} > 50~\mathrm{GeV}, ~|\eta_j| < 3,~|\eta_l| < 3,~ \Delta R_{ij} > 0.4,\\
  &M_{l_1l_2j_1j_2} > 300 ~\mathrm{GeV}, ~|M_{l_1l_2}-M_h| < 10~\mathrm{GeV}.
 \end{align}
  We list the cross sections for the $b\bar{b}\tau^{+}\tau^{-}$ and $b\bar{b}b\bar{b}$ final states after all cuts in Table \ref{table:s/b_sm}. After all cuts the cross section remains about $0.5~\rm{fb}$ ($13~\rm{fb}$)  for $b\bar{b}h$ with $h\to b\bar{b}$ at $14~ \mathrm{TeV}$ ($100~\mathrm{TeV}$), while the background is still large especially for $bbjj$ channel. The cut efficiency for the background process is about two orders of magnitude larger than that signal process, whereas the efficiency ratio for backgrounds can reach four orders at both 14~$\mathrm{TeV}$ and 100~$\mathrm{TeV}$. The $b\bar{b}h$ associated production with $h\to b\bar{b}$ is hard to be detected. For $b\bar{b}h$ with $h\to \tau^+\tau^-$, it is possible to do the study at a 100 $\mathrm{TeV}$ collider and the corresponding significance can reach 6.72 (12.3) with $L=300~\mathrm{fb}^{-1}$ ($L=1000~\mathrm{fb}^{-1}$).
\begin{table}[h!]
  \centering  \caption{The cross sections of the signal, background processes and the significances at LHC $14~\mathrm{TeV}$, $100~\mathrm{TeV}$ after all cuts.}\label{S/B}\label{table:s/b_sm}
  \begin{tabular}{|l|c|c|c|c|}
  \hline
  \multirow{2}*{} & \multicolumn{2}{c|}{$pp \to b\bar{b}h \to b\bar{b}b\bar{b}$}  & \multicolumn{2}{c|}{$pp \to b\bar{b}h \to b\bar{b}\tau^+\tau^-$} \\\cline{2-5}
   & 14~TeV & 100~TeV & 14~TeV &  100~TeV\\\hline
  $\sigma~(\mathrm{fb})$ & 0.46  & 13.4 & $5.7\times10^{-3}$ & 0.19\\
  $\sigma(BG)~(\mathrm{fb})$ & 1.0$\times10^4$  & $2.4\times 10^5$ &  $1.2\times10^{-2}$ &0.24 \\
  $S/\sqrt{B}$ with L=300~$\mathrm{fb}^{-1}$ & 0.08  & 0.47 & 0.90  &6.72 \\
  $S/\sqrt{B}$ with L=1000~$\mathrm{fb}^{-1}$ & 0.15  & 0.86 & 1.65  &12.3 \\
  \hline
\end{tabular}
\end{table}
 \subsection{Exotic Higgs results}
In this section, we investigate the associated production of $b\bar{b}$ and the heavy neutral Higgs (H or A) at hadron colliders. The parameters are set as Eq.\eqref{parameter1}, as well as  $m_{H}=m_A=500~\mathrm{GeV}$ and $\tan\beta=10$ for the typical examples.
\subsubsection{$b\bar{b}H$ and $b\bar{b}A$ production with $H~(A)\to b\bar{b}$}
\begin{figure}[!ht]
\begin{center}
\subfigure[]{
\includegraphics[scale=0.25]{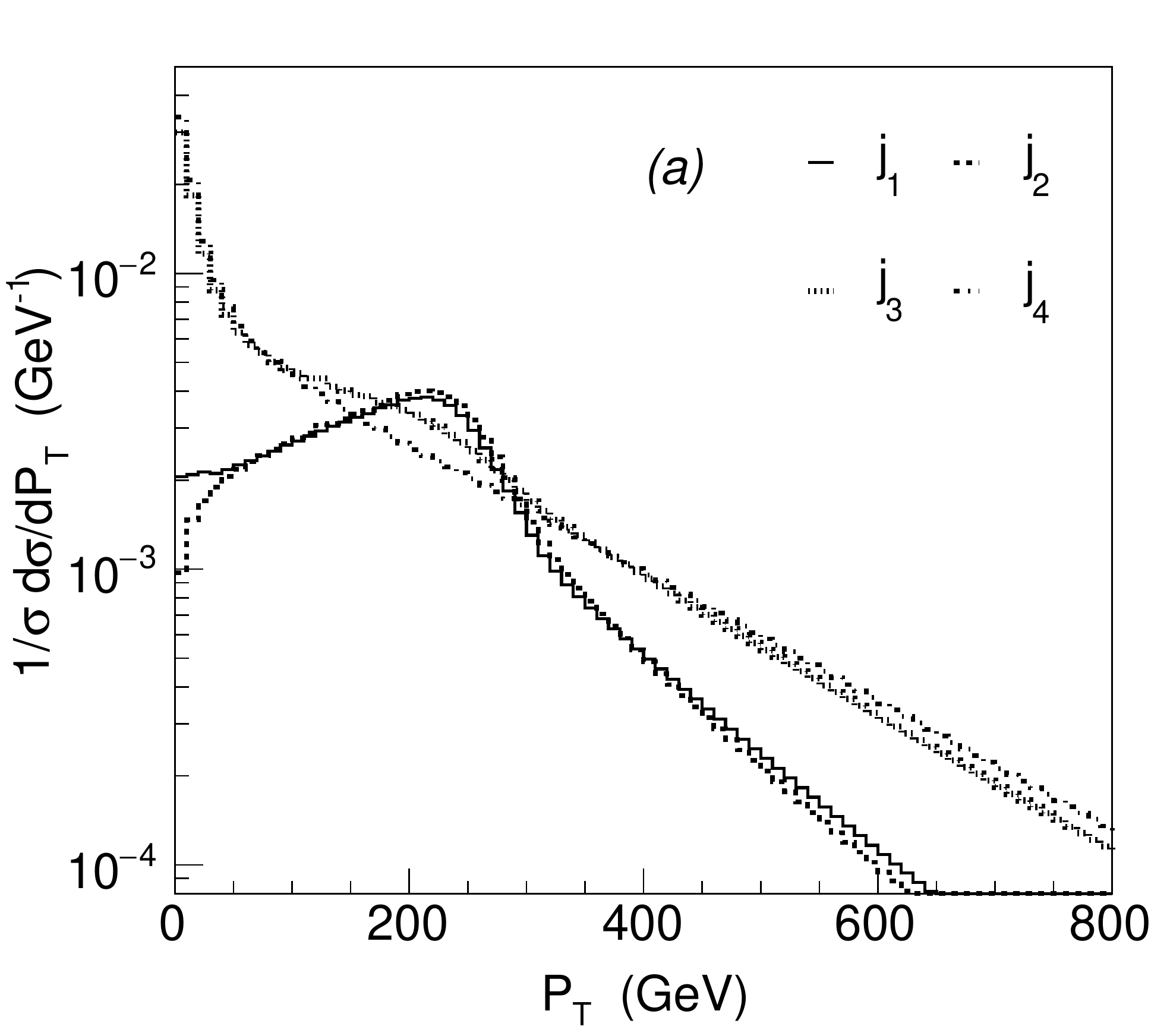}
}
\subfigure[]{
\includegraphics[scale=0.25]{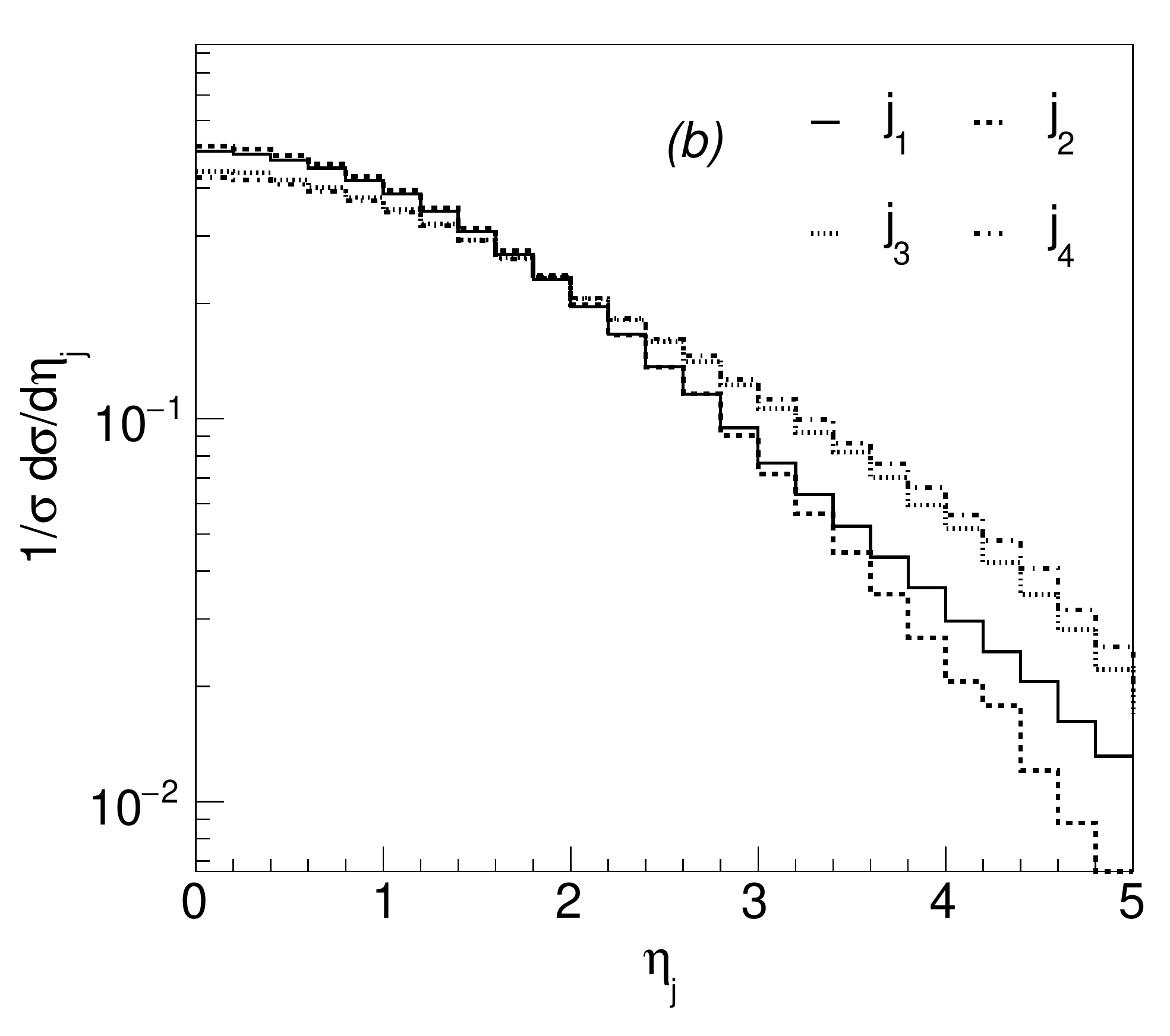}
}
\subfigure[]{
\includegraphics[scale=0.25]{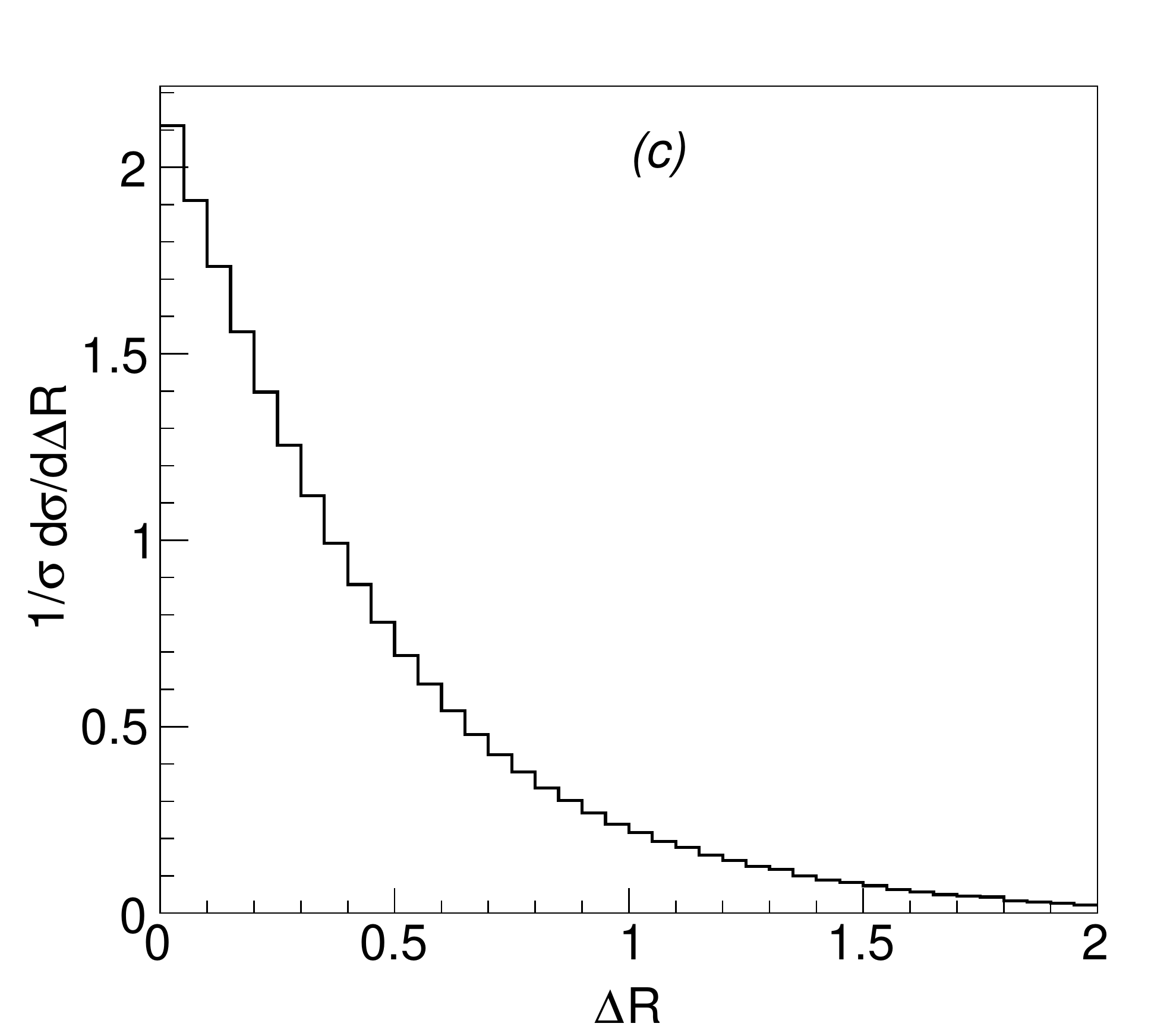}
}
\caption{(a) The normalized differential distributions with the transverse momentum of the jets ($p_T^{j_1}>p_T^{j_2}$
   and $p_T^{j_3}>p_T^{j_4}$)
   in the process of $pp \to b\bar{b}H \to b\bar{b}b\bar{b}$ for $M_H=500~\mathrm{GeV}$ at LHC $14~\mathrm{TeV}$. (b) The same as (a) with rapidity distributions. (c) The minimal angular separation distributions between jets.}\label{fig:4b_distrbution}
\end{center}
\end{figure}

 For the $b\bar{b}H$ production with $H\to b\bar{b}$, we study the transverse momentum distribution, rapidity distribution and minimal angular separation of jets $\Delta R$ distribution, where $\Delta R=min(\Delta R_{ij})$. The corresponding results are shown in Fig.\ref{fig:4b_distrbution}~(a), (b) and (c) respectively.
In order to improve the signal-background ratio, we apply the basic acceptance cuts,
\begin{align}
 ~~cut~I: ~~~~~ p_T^{j} &> 20~\mathrm{GeV}, ~|\eta_j| < 3,~ \Delta R > 0.4.
\end{align}

\begin{figure}[!ht]
\begin{center}
\subfigure[]{
\includegraphics[scale=0.35]{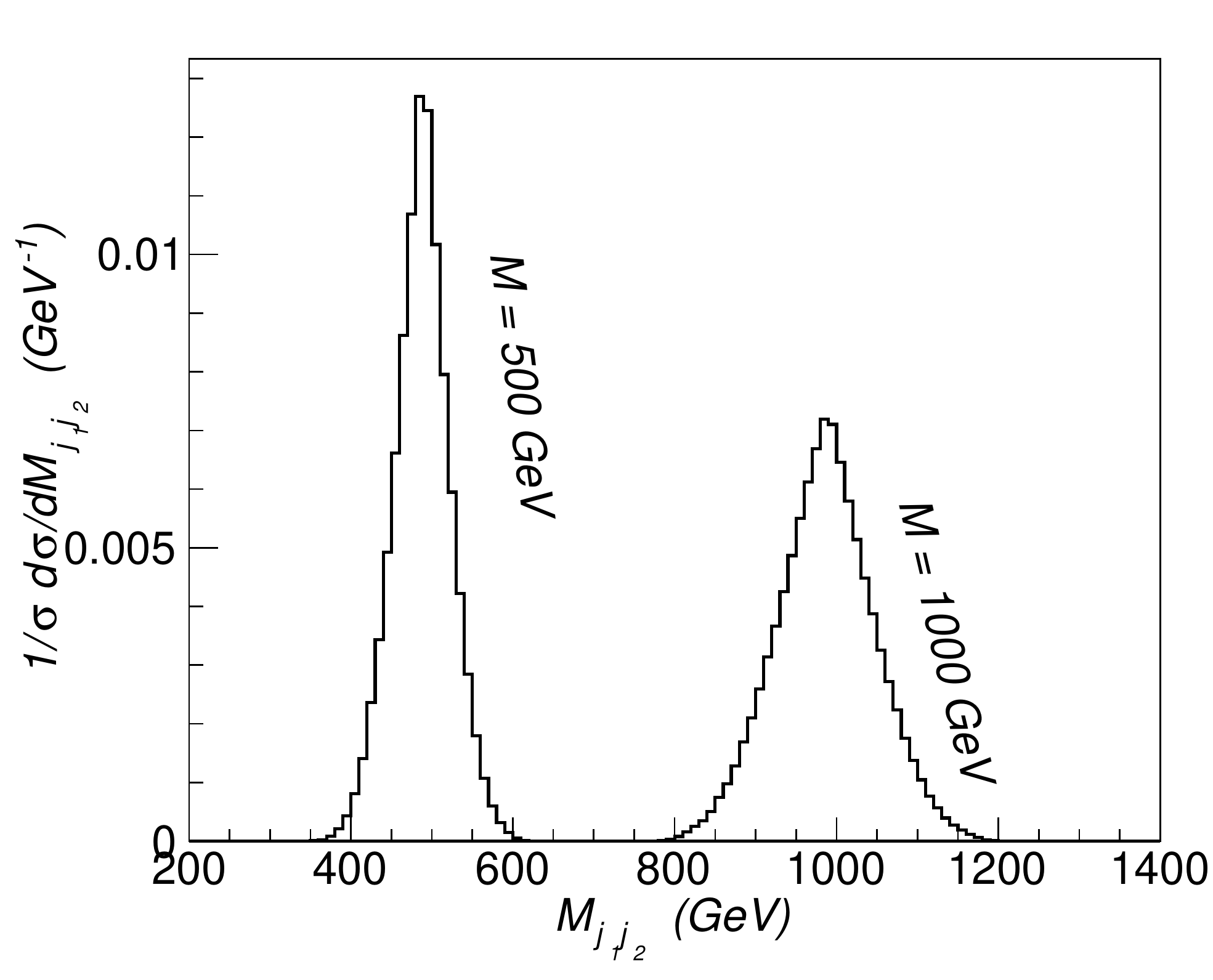}
}
\subfigure[]{
\includegraphics[scale=0.35]{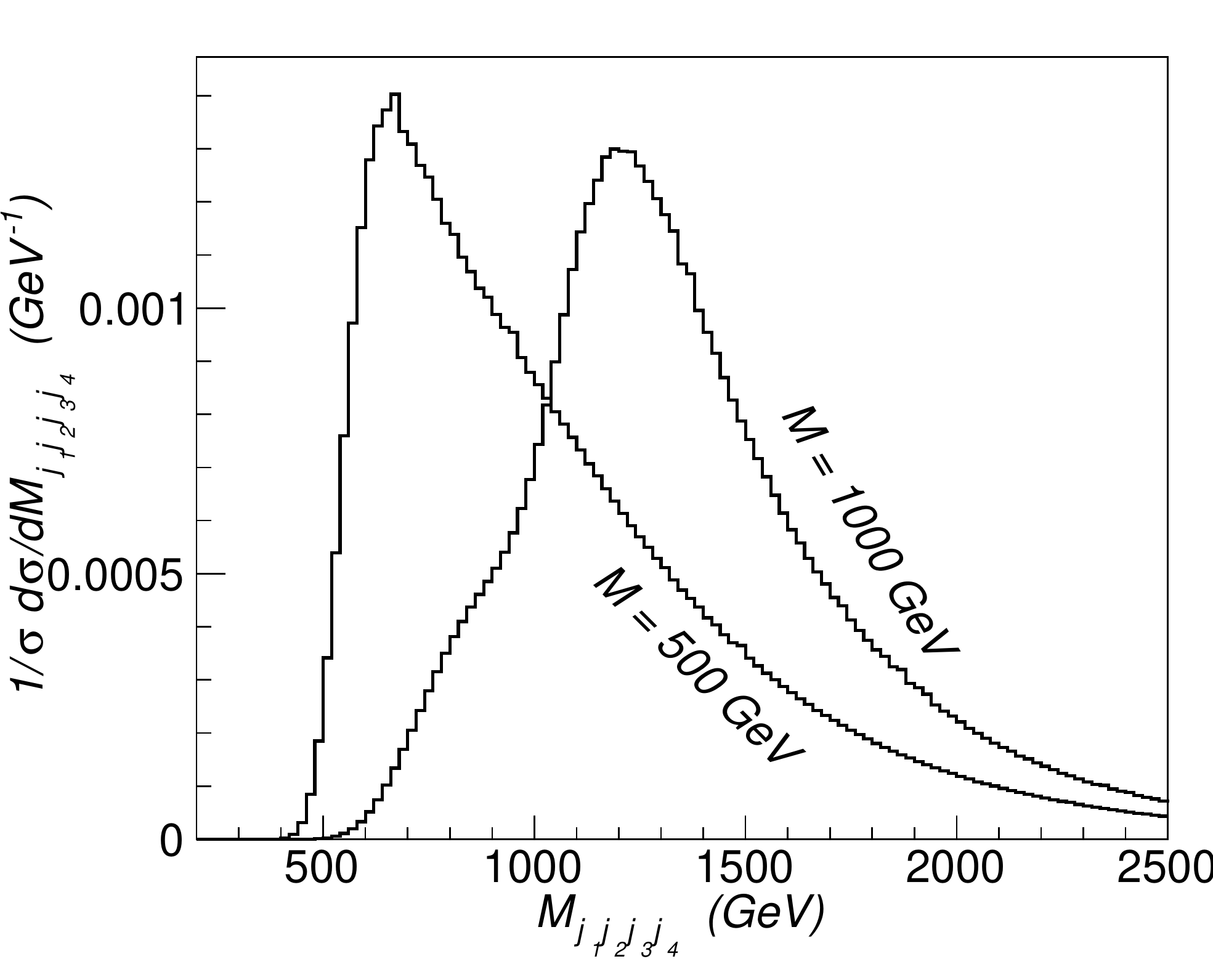}
}
\caption{(a) The distributions of $ 1/ \sigma (d\sigma/dM_{ij})$ with respect to the $j_1,j_2$ invariant mass with $M=500 ~\mathrm{GeV}$, $1000~\mathrm{GeV}$ for the process of $pp \to b\bar{b}H \to b\bar{b}b\bar{b}$ at the $14 ~\mathrm{TeV}$ LHC. (b) The same as (a) with $1/\sigma (d\sigma/M_{j_1j_2j_3j_4})$ distribution.} \label{fig:4b_Mjj}
\end{center}
\end{figure}

We also investigate the distribution of $1/\sigma (d\sigma/M_{j_1j_2j_3j_4})$ and the invariant mass distribution of $1/\sigma (d\sigma/dM_{j_1j_2})$ for $pp\to b\bar{b}H\to b\bar{b}b\bar{b}$ process. The results are demonstrated in Fig.\ref{fig:4b_Mjj}. Compared with the backgrounds, there is a clear resonance peak with the reconstructed invariant mass of $j_1,j_2$, thus it will be an effective cut to search for the exotic Higgs boson.
Because the Higgs mass is heavier than the relevant SM particles, to further purify the signal, we adopt the following cut,
\begin{align}
~~cut~II: ~~~~~ M_{j_1j_2j_3j_4} > 1.1 M_H,~~~|M_{j_1j_2}-M_H| < 0.1M_H.
\end{align}

In order to reduce $bbjj$ background, events with at least three identified b-tagged jets are selected for our analysis, i.e. $cut~IV: \mathrm{three~b~jet~tagging}$. The b-tagging efficiency is suppose to 60\% for the signal process for the simple simulation. The similar approach can be apply to $b\bar{b}A$ production with $A\to b\bar{b}$.

\begin{figure}[!ht]
\begin{center}
\subfigure[]{
\includegraphics[scale=0.35]{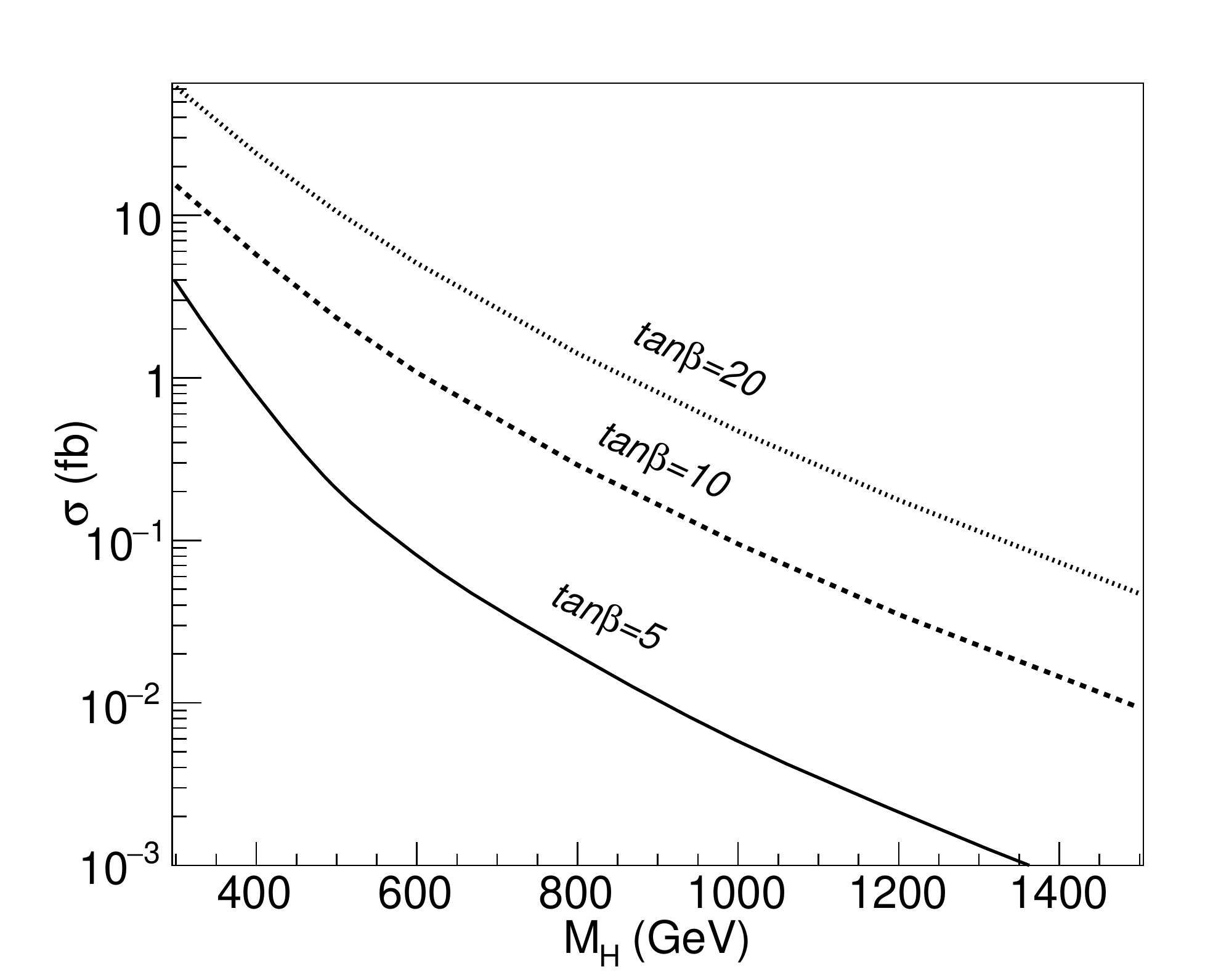}
}
\subfigure[]{
\includegraphics[scale=0.35]{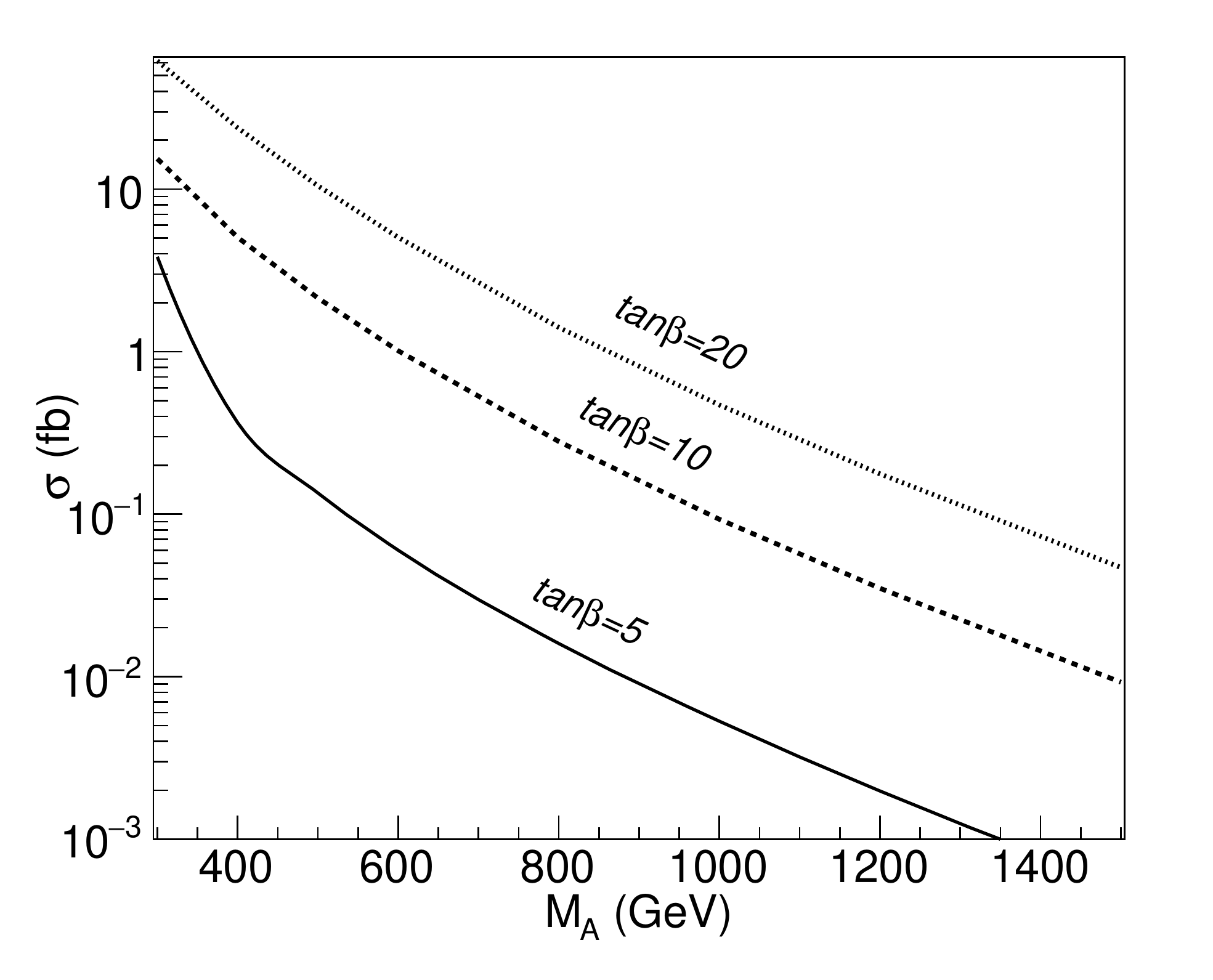}
}
\caption{The cross sections as a function of $M_{H/A}$ at the LHC for signal processes (a) $pp \to b\bar{b}H \to b\bar{b}b\bar{b}$ and (b) $pp \to b\bar{b}A \to b\bar{b}b\bar{b}$ with $\sqrt{s}=14~\mathrm{TeV}$ after all cuts.}
  \label{fig:4b_cs_aftercut}
\end{center}
\end{figure}

\begin{figure}[!ht]
\begin{center}
\subfigure[]{
\includegraphics[scale=0.35]{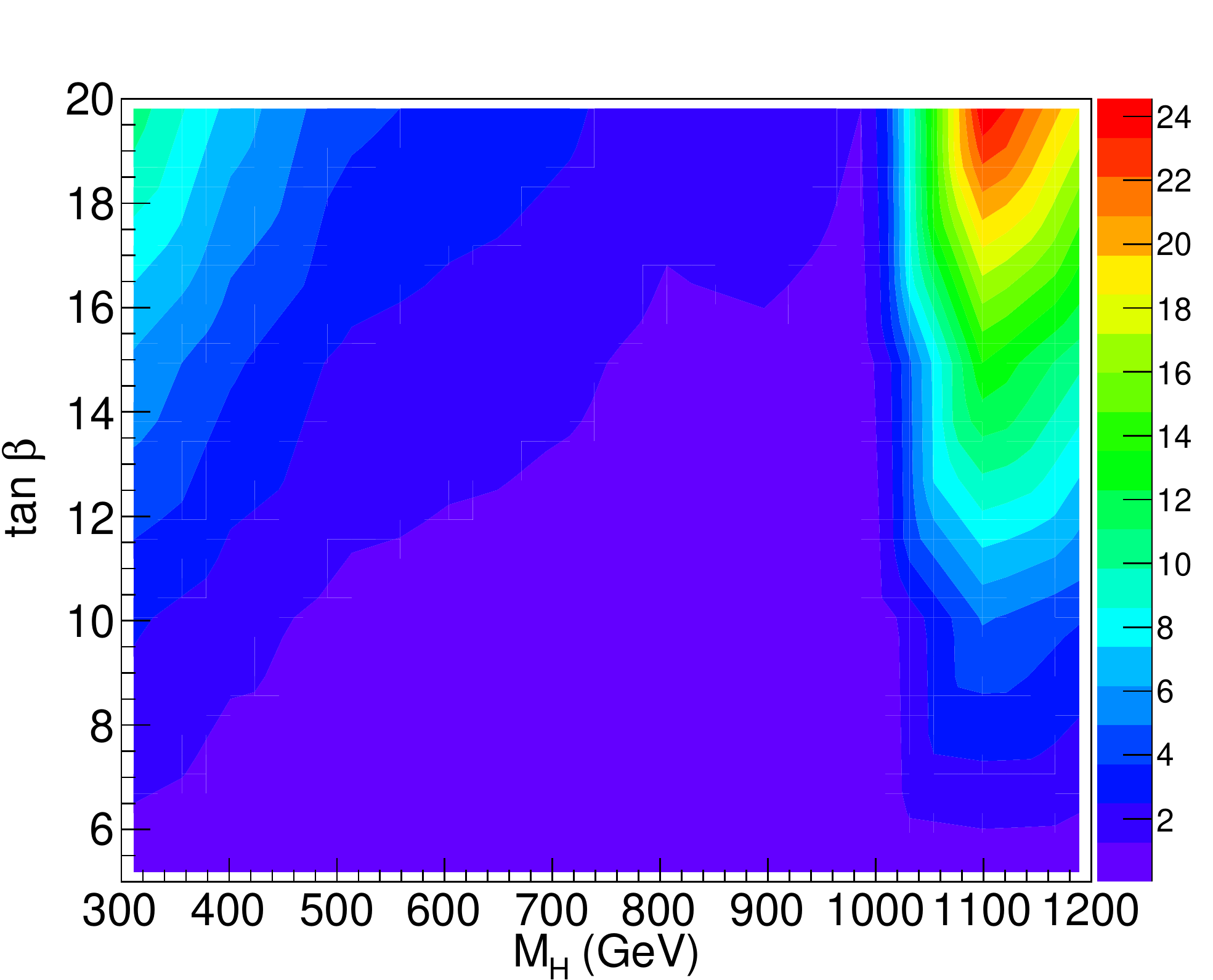}
}
\subfigure[]{
\includegraphics[scale=0.35]{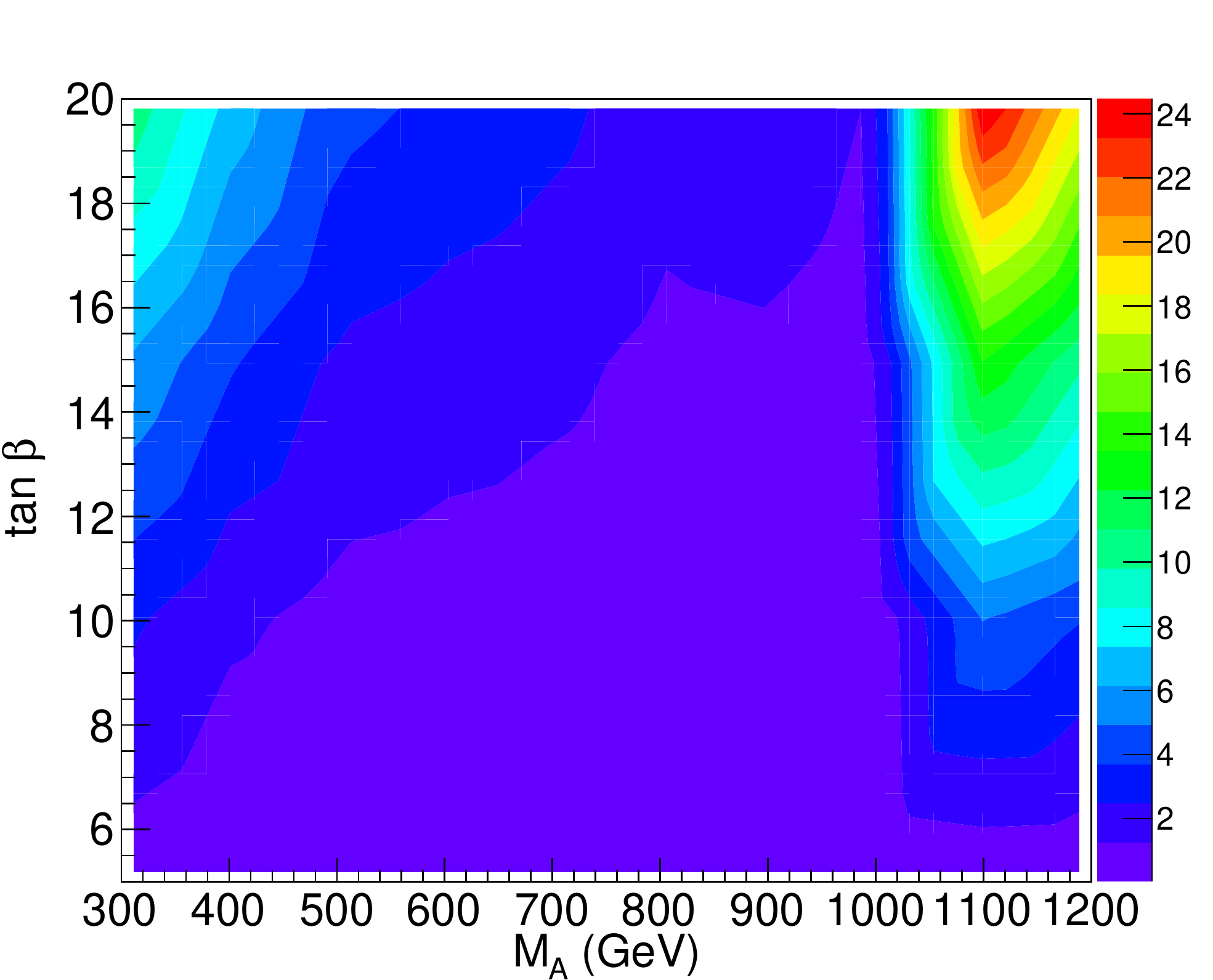}
}
\caption{(a) The significance as a function of $M_{H/A}$ and $\tan~\beta$ for process $pp \to b\bar{b}H \to b\bar{b}b\bar{b}$ at LHC $14~\mathrm{TeV}$ with the luminosity of 300 $\rm{fb}^{-1}$. (b) The same as (a) for $pp \to b\bar{b}A \to b\bar{b}b\bar{b}$ process.}
  \label{fig:4b_m_teba_s/b}
\end{center}
\end{figure}

The cross sections of $b\bar{b}H$ and $b\bar{b}A$ production with $H~(A)\to b\bar{b}$ at LHC 14 TeV after all cuts are displayed as a function of Higgs boson mass in Fig.\ref{fig:4b_cs_aftercut}.
The significance distributions in the $M_{H/A}-\tan\beta$ plane for $b\bar{b}H$ and $b\bar{b}A$ production with $H~(A)\to b\bar{b}$ at LHC 14 TeV with the integrated luminosity of 300 $\rm{fb}^{-1}$ are plotted in Fig.\ref{fig:4b_m_teba_s/b}. The result shows that corresponding to the same Higgs boson mass, the signal process will be easier to be detected at the large $\tan\beta$ region for the cross section is increasing with $\tan\beta$. With $m_{H/A}<1000~\rm{TeV}$, the signal process could be studied only in the region of small Higgs mass as well as $\tan\beta>15$. Our strategies are most effective in the mass region around 1100 GeV at the LHC, while the parameter space is difficult to study through the $b \bar{b}b \bar{b}$ final state with $600~\rm{GeV}<m_{H/A}<1000~\rm{GeV}$ and $\tan\beta<20$.
\subsubsection{$b\bar{b}H$ and $b\bar{b}A$ production with $H\ (A)\rightarrow\tau^+\tau^-$}
\begin{figure}[!ht]
\begin{center}
\subfigure[]{
\includegraphics[scale=0.25]{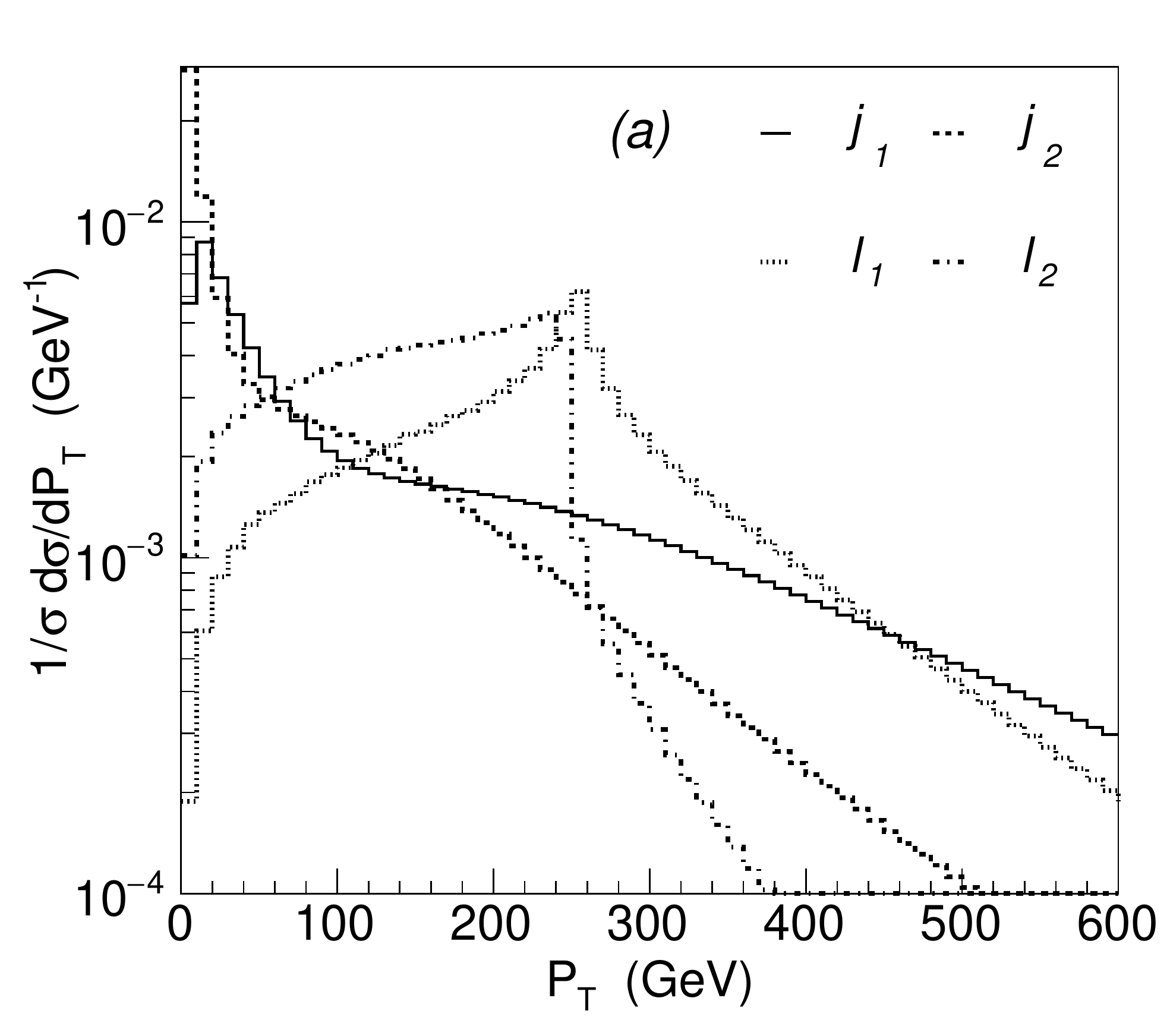}
}
\subfigure[]{
\includegraphics[scale=0.25]{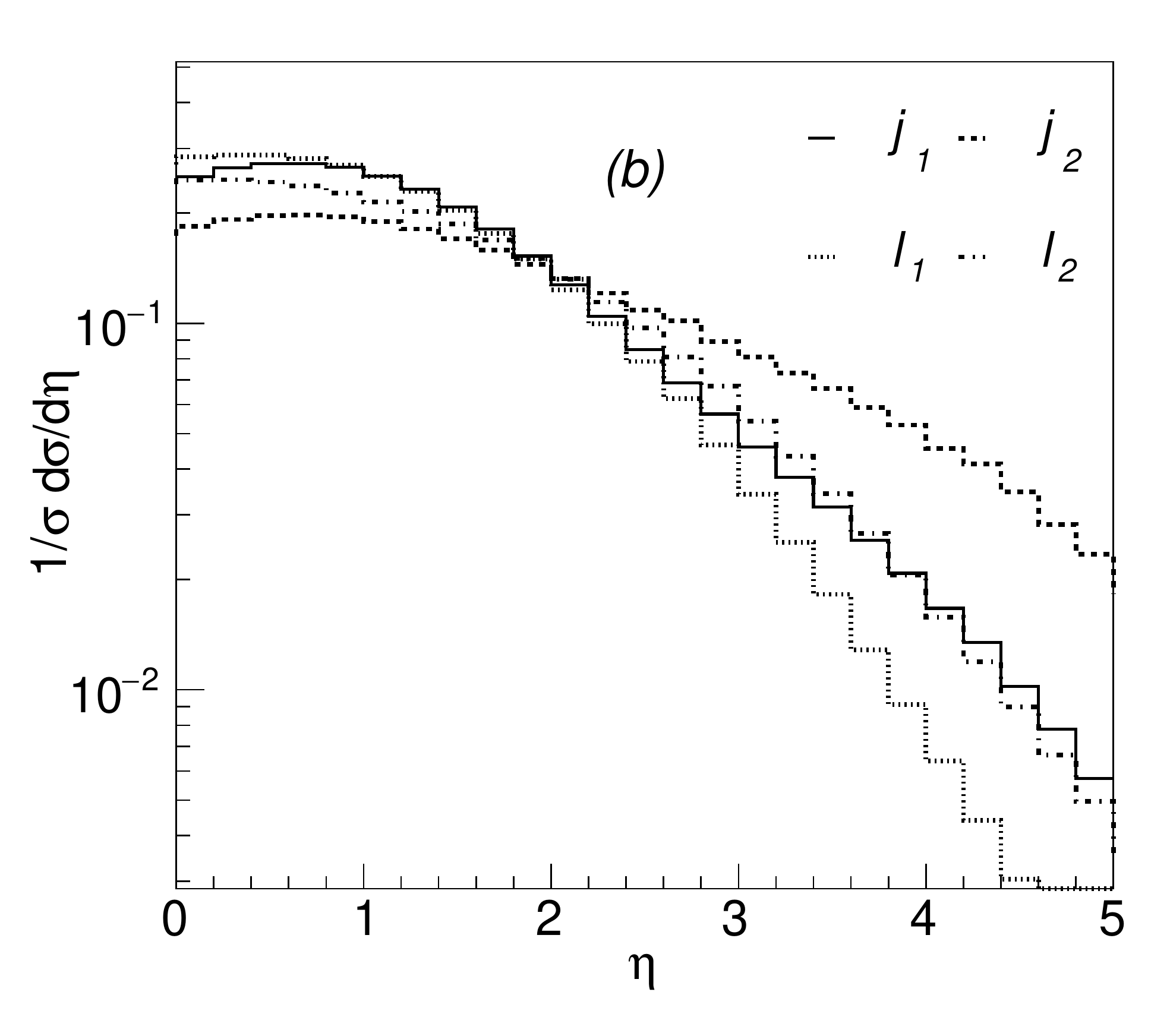}
}
\subfigure[]{
\includegraphics[scale=0.25]{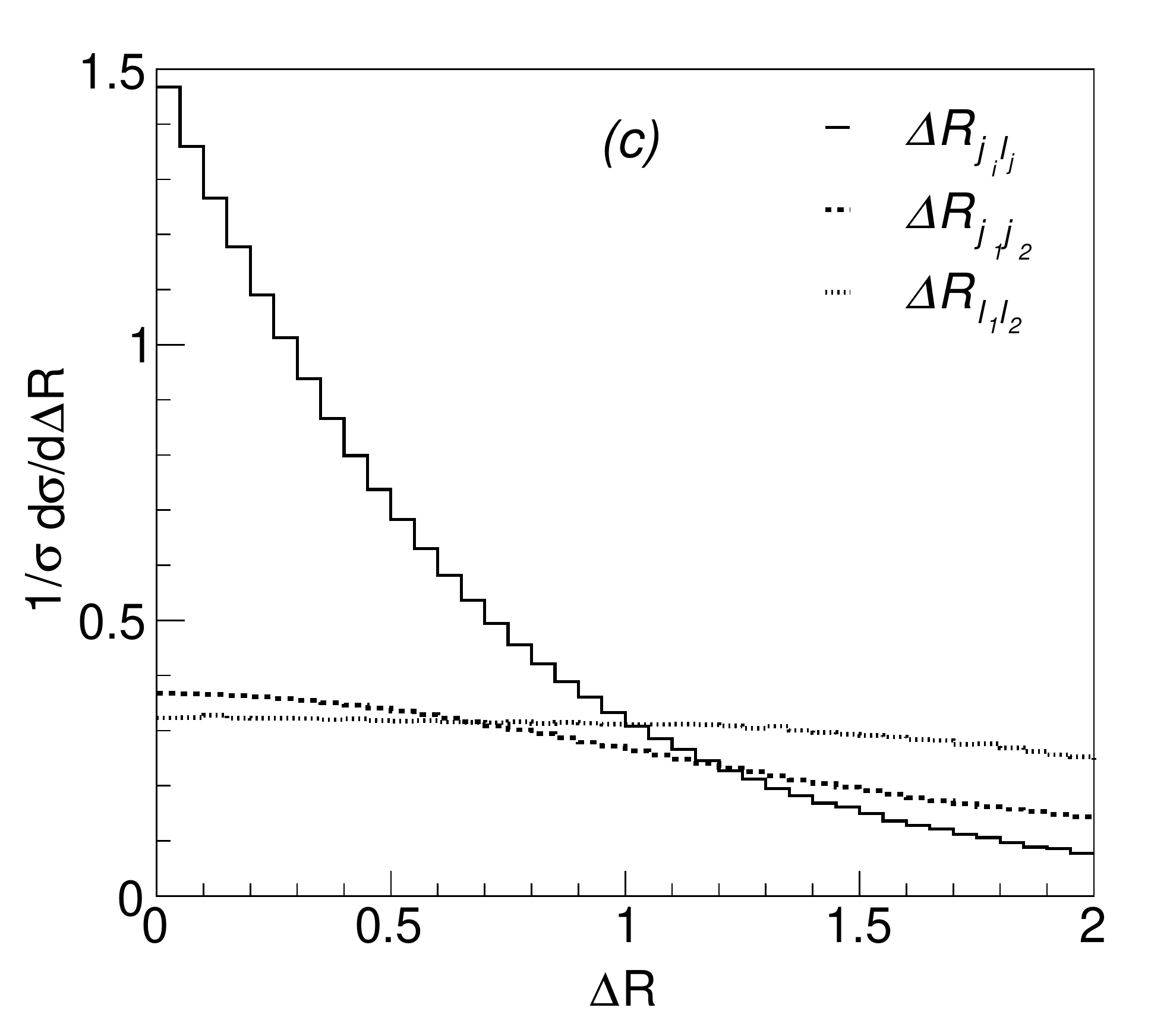}
}
\caption{(a) The normalized differential distributions with the transverse momentum of the jets ($p_T^{j_1}>p_T^{j_2}$)
   and charged lepton ($p_T^{l_1}>p_T^{l_2}$)
   in the process of $pp \to b\bar{b}H \to b\bar{b}\tau^+\tau^-$ $M_H=500~\mathrm{GeV}$ at LHC $14~\mathrm{TeV}$. (b) The same as (a) with rapidity distributions. (c) The minimal angular separation distributions between jets, jets and leptons and that between leptons.}\label{fig:bbtautau_distrbution}
\end{center}
\end{figure}

\begin{figure}[!ht]
\begin{center}
\subfigure[]{
\includegraphics[scale=0.35]{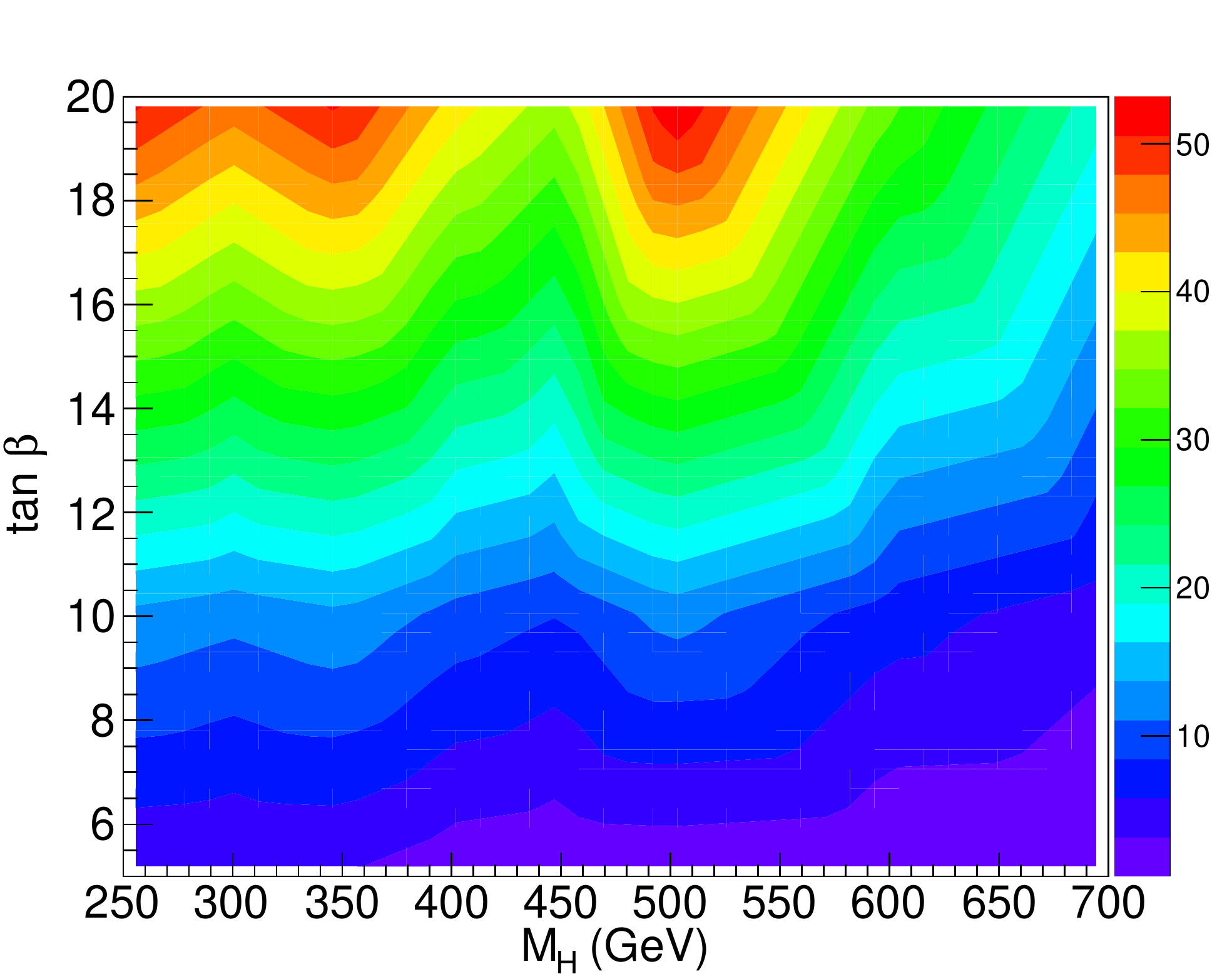}
}
\subfigure[]{
\includegraphics[scale=0.35]{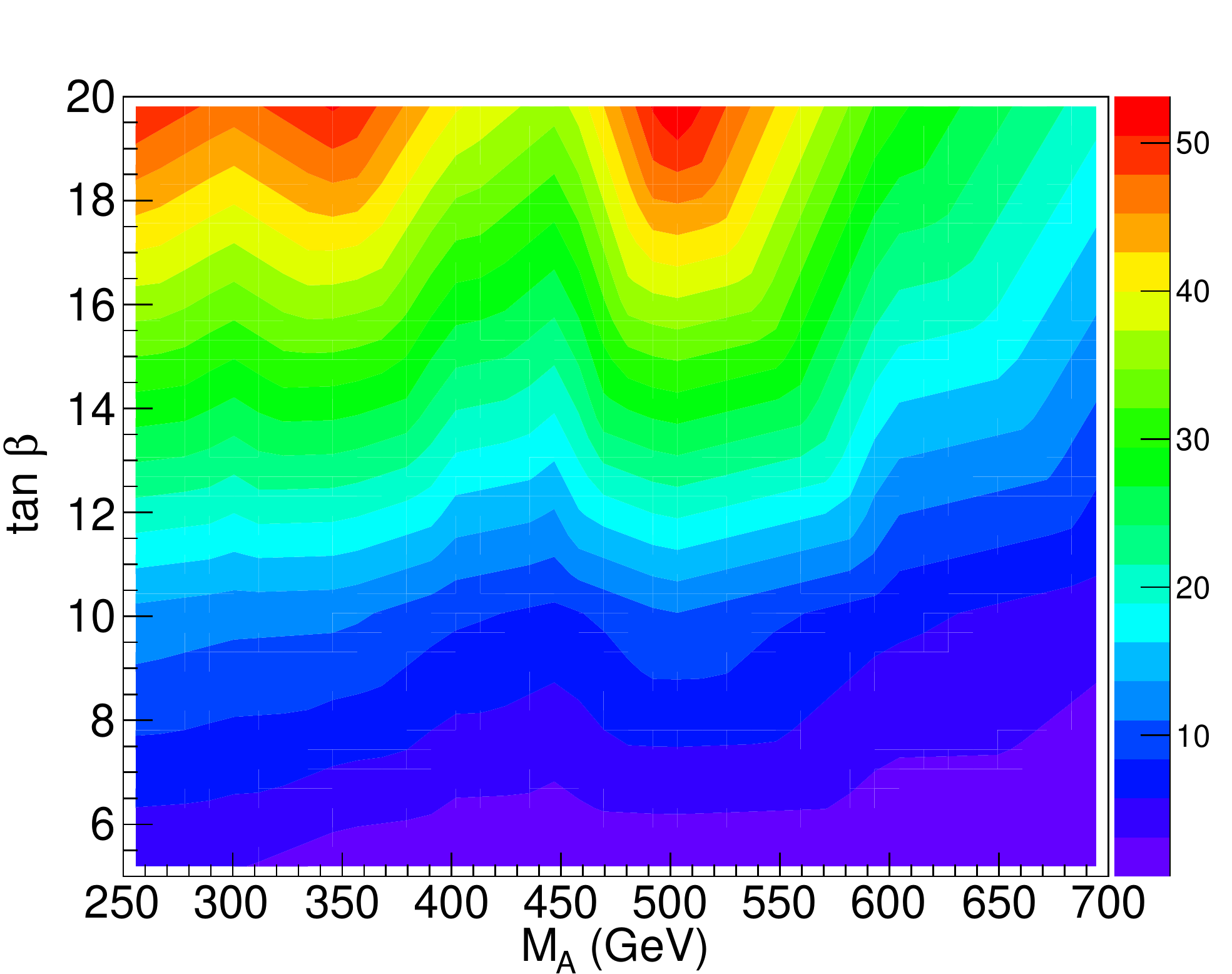}
}
\caption{(a) The significance as a function of $M_{H/A}$ and $\tan\beta$ for process $pp \to b\bar{b}H \to b\bar{b}\tau^+\tau^-$ at LHC $14~\mathrm{TeV}$ with the luminosity of 300  $\rm{fb}^{-1}$. (b) The same as (a) for $pp \to b\bar{b}A \to b\bar{b}\tau^+\tau^-$ process.}
  \label{fig:2b2tau_cs_m_s/b}
\end{center}
\end{figure}

%
The production rate of backgrounds with inclusive charged leptons is less than those with the multi-jets, therefore we investigate $b\bar{b}\tau^+\tau^-$ process at LHC. The $\tau$ leptons are tagged as $\tau$-jets with the hadronic decaying. The $\tau$-tagging efficiency depends on the $p_T$-distribution, while we set the tagging efficiency as 10\% for simplification according to the PGS detector simulation. The differential distributions with the transverse momentum and rapidity of the jets (charged leptons), and $\Delta R$ are shown in Fig.\ref{fig:bbtautau_distrbution}. For this signal process, the dominant SM background process is $pp\to Zbb$ process with $Z\to \tau^+\tau^-$.
Comparing the difference of kinematic distributions between the signal process and background, we employ the following cuts,
 \begin{align}
 &cut~I:   ~p_T^{j_2} > 20~\mathrm{GeV},~p_T^{l_2} > 20~\mathrm{GeV},~|\eta_{j}| < 3,~|\eta_{l}| < 3,~ \Delta R_{ij} > 0.4, \nonumber\\
 &cut~II:  ~p_T^{l_2} > 100~\mathrm{GeV}, \nonumber\\
 &cut~III: ~|M_{l_1l_2}-M_H| < 0.1M_H.
\end{align}
\begin{figure}[!ht]
\begin{center}
\subfigure[]{
\includegraphics[scale=0.35]{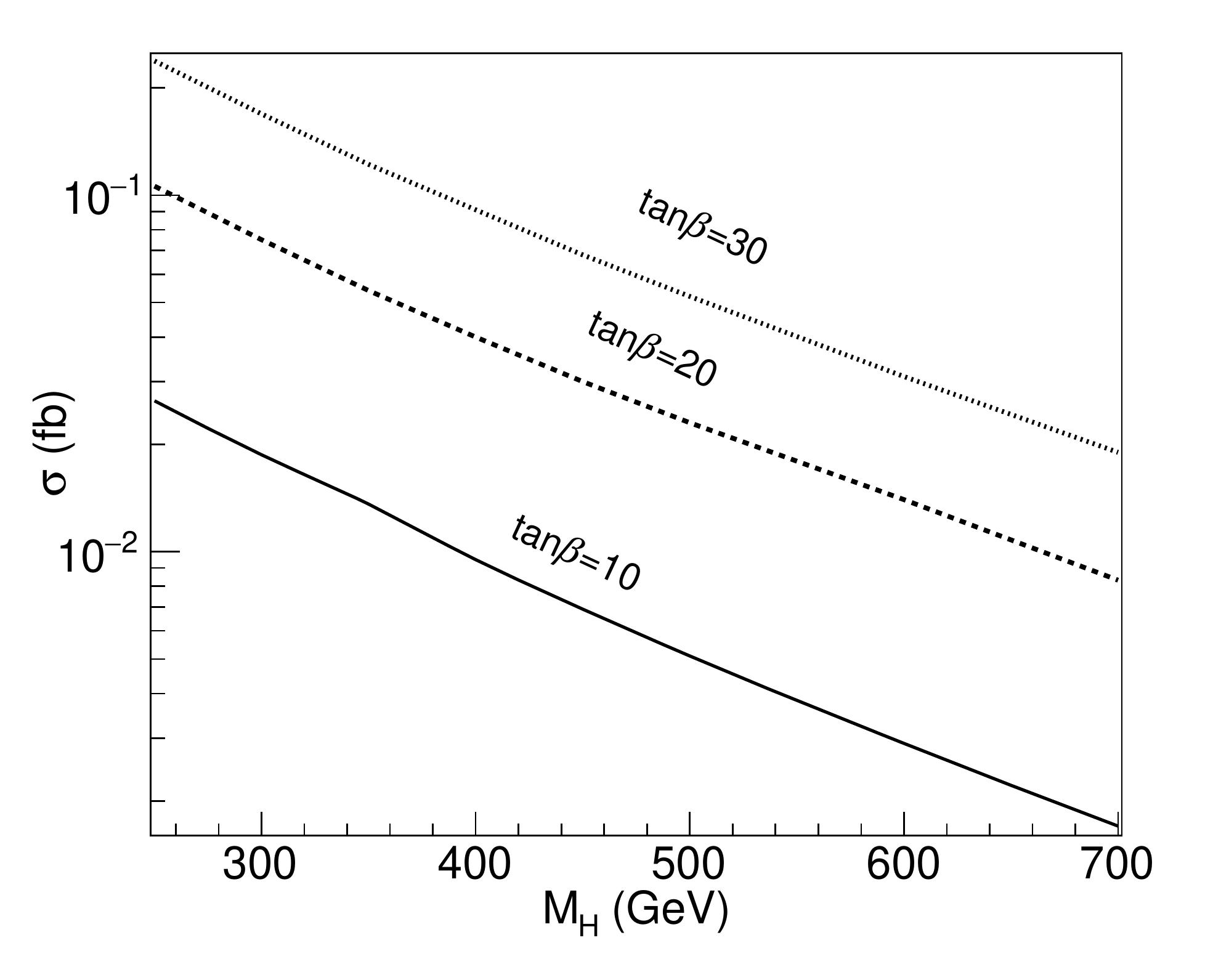}
}
\subfigure[]{
\includegraphics[scale=0.35]{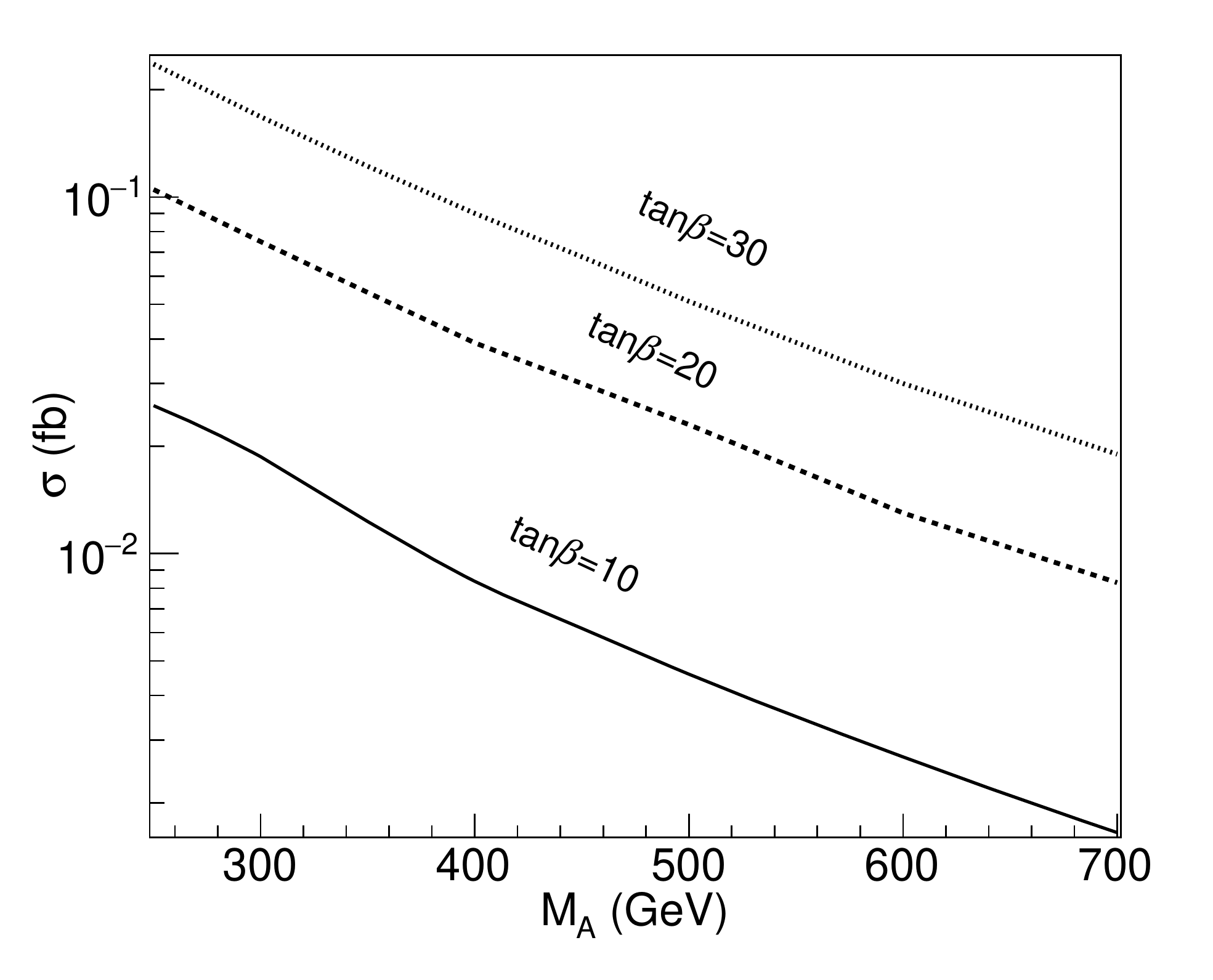}
}
\caption{The cross sections as a function of $M_{H/A}$ at LHC $14~\mathrm{TeV}$ for signal processes (a) $pp \to b\bar{b}H \to b\bar{b}\tau^+\tau^-$  and (b) $pp \to b\bar{b}A \to b\bar{b}\tau^+\tau^-$ after all cuts.}
  \label{fig:2b2tau_cs_aftercuts}
\end{center}
\end{figure}
In Fig.\ref{fig:2b2tau_cs_aftercuts}, we present the total cross sections of $b\bar{b}H$ and $b\bar{b}A$ production with $H~(A)\to \tau^+\tau^-$ after all cuts at LHC 14 TeV. Supposing the integral luminosity to be 300$~\rm{fb}^{-1}$ at $14~\mathrm{TeV}$, we display the significance distributions in the two dimension plots of $M_{H/A}-\tan\beta$ plane in Fig.\ref{fig:2b2tau_cs_m_s/b}. If the value of $\tan\beta<5$, it is difficult to observe $b\bar{b}H$ and $b\bar{b}A$ process with $300~\mathrm{GeV}<M_{H/A}<700~\mathrm{GeV}$ at LHC 14~$\mathrm{TeV}$. But in the region of $\tan\beta>5$, the cross section of signal is enhanced and the significance is larger than $3\sigma$ in most mass region.

\begin{figure}[!ht]
   \centering
   \includegraphics[scale=0.5]{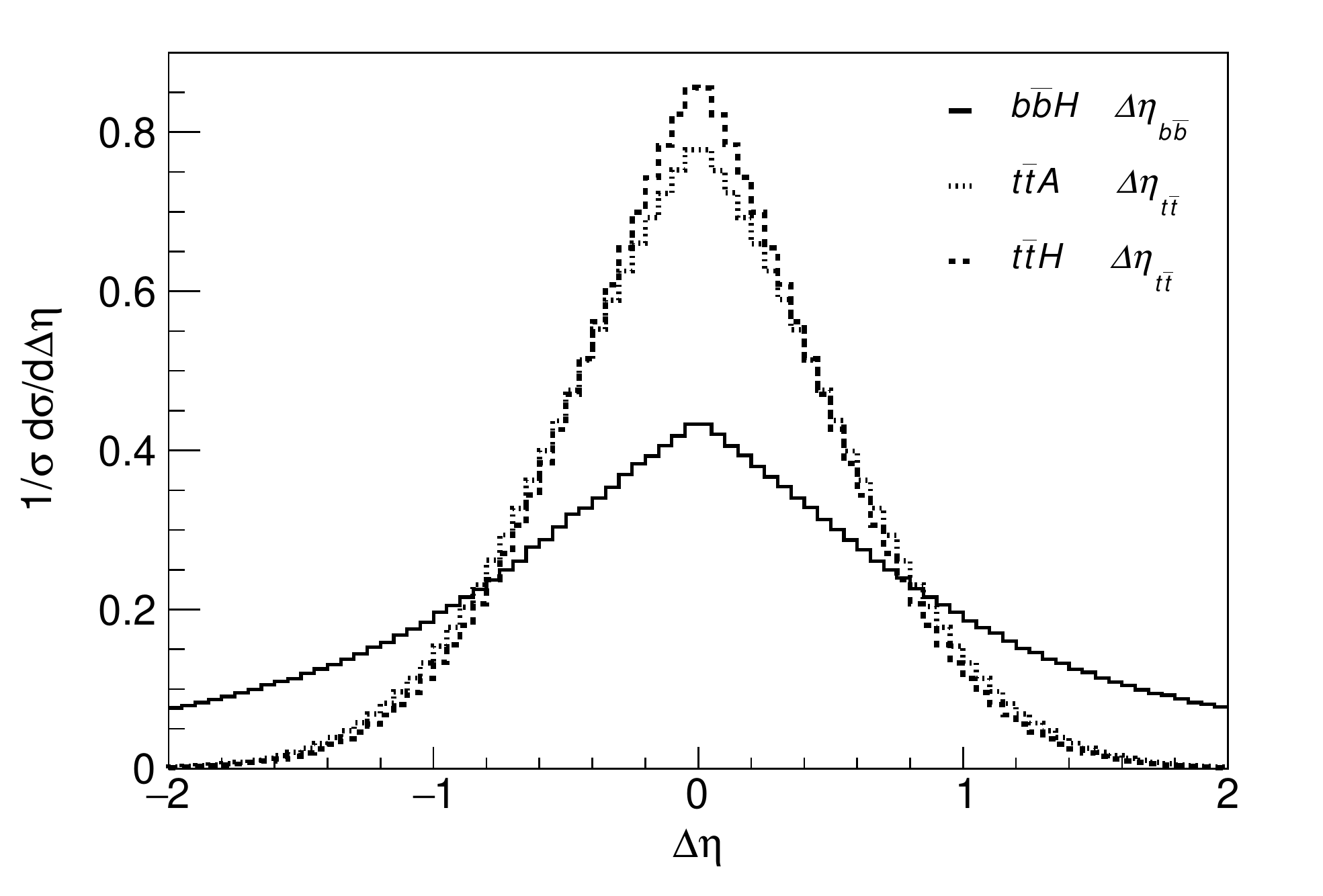}
   \caption{The Normalized differential distribution with the $\Delta \eta_{t\bar{t}/{b\bar{b}}}$ at LHC 14 TeV with $m_{H/A}=500$ GeV.}
   \label{fig:symmtrybbtt}
\end{figure}

In the case of the mass degeneration of H and A, the Yukawa interaction in Eq.\eqref{eq:lagrangian} is CP violating. Following the discussion of~\cite{Li:2017dyz}, the charge asymmetric term has a good discrimination power of the CP violating coupling.
The scalar and pseudoscalar interactions that contribute to the cross section of $b\bar{b}H$ production involve the factor $Tr(\slashed{p}_b\gamma_{\mu}\slashed{p}_{\bar{b}}\gamma_{\nu}\gamma_5)$, which is asymmetric in the interchange of $b$ and $\bar{b}$ and will affect the kinematics of the associated products of the bottom/anti-bottom quark. In Fig.\ref{fig:symmtrybbtt}, we show the differential distribution with $\Delta \eta$ for $b\bar{b}H/A$ and $t\bar{t}H/A$. $\Delta \eta$ is defined as the difference of rapidity of bottom$/$top and anti-bottom$/$anti-top quark from the associated production. $\Delta \eta_{b\bar{b}}$ is the same between $b\bar{b}H$ and $b\bar{b}A$ processes corresponding to solid curve. There are clear differences between scalar Higgs and pseudoscalar Higgs because of CP violating interactions in association with $t\bar{t}$. However the difference is tiny for the $b\bar{b}H$ and $b\bar{b}A$ due to the small CP violating interaction term proportional to the mass of bottom quark. Thus the CP properties are difficult to study via the Higgs boson associated production with the bottom quark pair.
\section{Summary}\label{Sec:Sum}

In this work, we study the production of $b\bar{b}\phi_k$ ($\phi_k=h,H,A$) at a hadron collider. For the SM Higgs (h), the $b\bar{b}h$ process is hard to be observed at LHC 14 TeV due to the large background, but for the exotic Heavy Higgs (H, A), the $b\bar{b}H$ and $b\bar{b}A$ production can be enhanced by $\tan\beta$. It is found that the cross section of $b\bar{b}H$ and $b\bar{b}A$ can be up to $200~\mathrm{fb}$ for $m_{H/A}=500~\mathrm{GeV}$ and $\tan\beta=10$ at LHC 14~$\mathrm{TeV}$. After considering $b\bar{b}H$ and $b\bar{b}A$ production with the subsequent Higgs decay $H~(A)\to b\bar{b}$, the heavy Higgs production signal can be observed in the mass region $300~\mathrm{GeV} <m_{H/A}<400~\mathrm{GeV}$ and $1000~\mathrm{GeV} <m_{H/A}<1200~\mathrm{GeV}$ for $\tan\beta>8$. For the $b\bar{b}\tau^+\tau^-$ channel, its significance after all cuts is higher than that of $b\bar{b}b\bar{b}$, which implies the $\tau$-tagging may be another useful tool to search for heavy resonance production. We also investigate the rapidity difference distribution for $b\bar{b}H$ ($b\bar{b}A$) and $t\bar{t}H$ ($t\bar{t}A$) at LHC, and find that charge asymmetry is sensitive to the heavy quarks mass, and may be used to discriminate the different CP property related to Higgs-top coupling. Our work shows that $b\bar{b}b\bar{b}$ and $b\bar{b}\tau^+\tau^-$ processes will play an important role to search for the production of the exotic heavy Higgs or other heavy resonance particles at LHC and other future hadron colliders.
\section*{Acknowledgement}
This work is supported in part by the National Natural Science Foundation of China (NSFC) under grant Nos.11875179, 11775130, 11635009 and 11605075 and Natural Science Foundation of Shandong Province under grant Nos. ZR2017JL006 and ZR2017MA002.

\end{document}